\documentclass[aps,pra,reprint,superscriptaddress,footinbib,longbibliography]{revtex4-2}

\usepackage[english]{babel}
\usepackage[utf8]{inputenc}
\usepackage[colorinlistoftodos, color=green!40, prependcaption]{todonotes}

\usepackage{amssymb}
\usepackage{amsmath}
\usepackage{graphicx}
\usepackage{amsthm}
\usepackage{mathtools}
\usepackage{xcolor}
\usepackage{graphicx}
\usepackage[left=23mm,right=13mm,top=35mm,columnsep=15pt]{geometry} 
\usepackage{adjustbox}
\usepackage{placeins}
\usepackage[T1]{fontenc}
\usepackage{xfrac}

\usepackage[unicode]{hyperref}
\hypersetup{
   unicode=true,          
   plainpages=false,
   colorlinks=true,       
   citecolor=blue,        
}

\def\BdG {\mathrm{BdG}}

\def\bf {\mathrm{bf}}
\def\ef{\varepsilon_\mathrm{F}}
\def\lf{l_\mathrm{F}}
\def\kf{k_\mathrm{F}}
\def\vf{v_\mathrm{F}}

\def\dd{\mathrm{d}}
\def\s{\mathrm{s}}
\def\d{\mathrm{d}}
\def\t{\mathrm{t}}
\def\pt{p_\mathrm{t}}
\def\r{\mathbf{r}}

\begin{document}
\title{Swallow-tail dispersions of moving solitons in a two-dimensional fermionic superfluid}
\author{Jan Major}
\email{jan.j.major@gmail.com}
\affiliation{Dodd-Walls Centre for Photonic and Quantum
Technologies, Centre for Theoretical Chemistry and
Physics, New Zealand Institute for Advanced Study,
Massey University, Private Bag 102904, North Shore,
Auckland 0745, New Zealand}
\affiliation{Institute of Theoretical Physics, Wrocław University of Science and Technology, 50-370 Wrocław, Poland}
\author{Joachim Brand}
\affiliation{Dodd-Walls Centre for Photonic and Quantum
Technologies, Centre for Theoretical Chemistry and
Physics, New Zealand Institute for Advanced Study,
Massey University, Private Bag 102904, North Shore,
Auckland 0745, New Zealand}

\date{\today}
\begin{abstract}
Soliton-like localised wave solutions in a two-dimensional Fermi superfluid are studied by solving the Bogoliubov-de Gennes equations in the BCS regime of weak pairing interactions. The dispersion relations of these solitons are found to exhibit a peculiar swallow-tail shape, with cusps and multiple branches. The effective mass of the solitons is found to diverge and change sign at the cusp. This behavior is in contrast to the smooth dispersion relations and negative effective masses of solitons in the three-dimensional Fermi superfluid. The swallow-tail dispersion relations are shown to be a consequence of  counterflow of the superfluid and  sign-changing contributions to the superfluid current from different transverse momenta in the Bogoliubov-de Gennes formalism. The results are relevant for the understanding of solitonic excitations in two-dimensional Fermi superfluids, such as ultracold atomic gases and high-temperature superconductors.
\end{abstract}

\maketitle

\section{Introduction}
Phenomena of superfluidity and superconductivity are known already for more than a hundred years \cite{Onnes1911,Kapitza1938,Allen1938}, yet they still inspire new research and provide new riddles to solve. The same can be said about solitons with an even longer research history \cite{Korteweg1895} and connections to transport and critical phenomena in a wide variety of systems: from DNA and polymers to cosmological models \cite{Muto1988,ZUREK1996}. Solitonic excitations in superfluids appear in form of dark solitons, which are characterised by an abrupt change of the complex phase of the superfluid order parameter \cite{Tsuzuki1971}. 
Dark solitons were confirmed and characterised experimentally in Bose-Einstein condensates \cite{Burger1999,Denschlag2000,Becker2008}, but in fermionic superfluids, including superconductors, their existence is still a matter of debate. Only observations of transient phenomena attributed to solitons have been reported so far \cite{Ku2016}. Theoretical studies of solitons in fermionic superfluids have been conducted in one dimension \cite{Shamailov2016,Efimkin2015,Zou2016,Dziarmaga2004c} and three dimensions \cite{Antezza2007,Liao11pr:FermiSolitons,Scott2011,Spuntarelli2011,Scott2012,Klimin2014,VanAlphen2019a}, but to the best of our knowledge, no studies of moving solitons in two-dimensional fermionic superfluids have been reported.

The thorough understanding of soliton properties and their connection to transport in fermionic superfluids (including superconductors) may be vital for progress in such fields as 
high-$T_c$ superconductivity \cite{Keimer2015} or in the analysis of 
novel topologically nontrivial superfluid systems \cite{Sato2017}.
This paper focuses on the case of a balanced s-wave superfluid in two dimensions in the crossover regime close to the Bardeen-Cooper-Schrieffer (BCS) limit. In this regime we observe some previously unseen behavior that is directly connected to the dynamic properties of solitons. For low binding energies, the dispersion relation undergoes a qualitative change, from a smooth inverted lobe into a swallow-tail shape with non-smooth cusps. From this change results a range of phenomena such as the coexistence of several solitonic solutions for the same velocity. We also find sign changes, zero crossings and divergence of the soliton's effective mass.

As the movement of solitons in many classes of systems is hindered by impurities, the ideal medium for studying soliton dynamics are ``clean'' systems of ultracold atoms \cite{Challis2007}. Apart from the lack of defects (unless intentionally included), those systems allow for unprecedented control over the parameters. This makes it possible, for example, to choose the  dimensionality, create external static and dynamic potentials, and generate artificial gauge fields such as spin-orbit coupling \cite{Bloch2012}. The most useful in the context of this paper is the possibility of changing the interparticle interaction strength via magnetic-field tunable Feshbach  resonances \cite{Chin2010}. This means that a range of models can be simulated: from the weakly-coupled BCS phase through the crossover to the Bose-Einstein condensate (BEC) of tightly bound fermionic pairs.

There are several theoretical descriptions of solitons in fermionic superfluids. In one dimensional superfluids, solitons have been studied both in full many body framework using an exactly-solvable model \cite{Shamailov2016} as well as using an analytically solvable mean field approximation \cite{Efimkin2015}, showing some significant differences.
In three dimensions mean field theory generally requires numerical solutions \cite{Antezza2007,Bulgac2011} although scaling relations provide some analytical insights for the unitary gas \cite{Liao11pr:FermiSolitons}. Our study aims at filling the gap for two-dimensional case and uses numerical solutions of the Bogoliubov-de Gennes (BdG) equations. We are focusing on the deep BCS regime near the BCS limit. The reason for that is twofold: It is known that the BdG approach fails in the BEC regime in two dimensions but is expected to be safer to use in weakly interacting regime \cite{Bertaina2011}. Moreover, we find very interesting phenomena for systems in the deep BCS regime.

In Sec.~\ref{sec:model} we describe the fermionic superfluid model studied in this paper as well as the mean-field BdG approach used to simulate its excitations. In Sec.~\ref{sec:scresults} we show results of the self-consistent calculations hosting solitonic excitations and examine their spatial profiles. 
The dispersion relations for these solitons  are presented in Sec.~\ref{soc:masses} along with calculations of their inertial and physical masses. In section~\ref{sec:counterflow} we focus on the
phenomenon of counterflow -- the fact that the sign of contributions to superfluid current may vary for different transverse momenta. Finally we conclude in \ref{sec:conclusion}.

Appendix~\ref{app:mu} contains a discussion of how superfluid flow affects the  chemical potential and App.~\ref{app:gb} discusses the change of the free energy of a soliton due to a change of reference frames.

\section{Model and method\label{sec:model}}
We are considering a balanced Fermi gas with contact $s$-wave interactions at zero temperature. The grand canonical Hamiltonian of such a system reads:
\begin{align}
 \mathcal{H}= \sum_\sigma\int \dd\r \left\{ \psi^\dagger_\sigma(\r)\hat{h}\psi_\sigma(\r) + g\psi^\dagger_\sigma(\r) \psi_{\bar{\sigma}}^\dagger(\r) \psi_{\bar{\sigma}}(\r) \psi_\sigma(\r) \right\},
\end{align}
where $\hat{h}=-\frac{\hbar^2}{2m}\nabla^2 -\mu$ is the one particle Hamiltonian, with $\mu$ the chemical potential and $m$ the fermion mass. 
$g$ is a renormalized interaction constant, $\sigma$ denotes the spin degrees of freedom ($\pm\sfrac{1}{2}$) and $\bar{\sigma}$ is opposite to $\sigma$.
A standard way of deriving a mean-field theory for an inhomogeneous superfluid system is the BdG approach \cite{DeGennes1999}. As it is a mean-field method, quantum correlations are neglected and it is expected to fail in low dimensions. In two-dimensional systems it is considered to be a good approximation in the BCS limit, while it is known to fail when approaching the BEC limit \cite{Bertaina2011}.

In this framework the system is described using the self-consistent BdG equation:
\begin{align}\label{eq:BdGmatrix}
    i\hbar\partial_t\left[\begin{array}{c}u(\r,t)\\ v(\r,t)\end{array}\right]= \left[\begin{array}{cc}\hat{h}&\Delta(\r,t)\\\Delta^*(\r,t)&-\hat{h}\end{array} \right]\left[\begin{array}{c}u(\r,t)\\ v(\r,t)\end{array}\right].
\end{align}
The $u(\r,t)$ and $v(\r,t)$ are quasi-particle amplitudes which act as weights for the approximated many body wave function.
The superconducting order parameter $\Delta(\r,t)$ has to be determined self-consistently from a complete and orthogonal set of solutions for the quasi-particle amplitudes \cite{Challis2007}.

We will further restrict ourselves to two-dimensional systems and consider solutions which are: 1) inhomogeneous only in the $x$ direction, 2) are moving with velocity $v_s$ in direction $x$. Because translational invariance in the transverse $y$ direction is maintained, the BdG matrix splits into disconnected blocks according to different transverse momentum channels. We
rewrite it in the frame co-moving with the soliton \cite{Liao11pr:FermiSolitons}. Eventually we can obtain a time independent version of Eq.~\eqref{eq:BdGmatrix} for each transverse momentum $\pt$:
\begin{align}
H^{(\pt)}_\BdG\left[\begin{array}{c}u_\nu^{(\pt)}(x)\\ v_\nu^{(\pt)}(x)\end{array}\right]=
 \varepsilon_\nu^{(\pt)}\left[\begin{array}{c}u_\nu^{(\pt)}(x)\\ v_\nu^{(\pt)}(x)\end{array}\right] ,
\end{align}
where $\nu$ is the index of the eigenstate and $\varepsilon_\nu^{(\pt)}$ is the corresponding eigenenergy. The BdG matrix for each $\pt$ is given by
\begin{align} \label{eq:Hpt}
    H^{(\pt)}_\BdG =\left[\begin{array}{cc}\hat{h}_x+\frac{\pt^2}{2m}&\Delta(x)e^{2imv_\s x/\hbar}\\\Delta^*(x)e^{-2imv_\s x/\hbar}&-\hat{h}_x+\frac{\pt^2}{2m}\end{array} \right],
\end{align}
where $\hat{h}_x = -\frac{\hbar^2}{2m} \partial_x^2 - \mu$ is a differential operator acting on $x$ only.
The quasiparticle amplitudes $u_\nu^{(\pt)}(x)$ and $v_\nu^{(\pt)}(x)$ are eigenstates of the respective $H^{(\pt)}_\BdG$ matrix. $\Delta(x)$ has to be calculated self-consistently as
\begin{align}\label{eq:deltaPT}
    \Delta(x)=-g\sum_\nu^{\prime}\sum_{\pt} u_\nu^{(\pt)}(x){v_\nu^{(\pt)}}^*(x),
\end{align}
combining contributions from all transverse momentum channels.
The particle number density is given by
\begin{align} \label{eq:n}
  n(x)=\sum_\nu^{\prime}\sum_{\pt}|v_\nu^{(\pt)}(x)|^2,
\end{align}
where the primed sum is done over states with positive eigenenergies ($\sum'_\nu\equiv\sum_{\nu: \varepsilon_\nu^{(\pt)}>0}$). 

Solutions resembling dark solitons on an infinite domain are expected to have a phase jump in $\Delta(x)$, $\phi_\mathrm{s} = \arg[\Delta(\infty) / \Delta(-\infty)]$, which is a signature of the soliton and is expected to depend on $v_\mathrm{s}$ \cite{Tsuzuki1971,Liao11pr:FermiSolitons,Scott2011}. In addition, the BdG matrix of Eq.~\eqref{eq:Hpt} has a phase gradient in the off-diagonal terms, originating from a frame transformation into the soliton's rest frame. In order to solve the BdG equations with periodic (or anti-periodic) boundary conditions, it is therefore convenient to use a unitary transformation that, in combination with a convenient boundary condition, allows us to keep only the residual backflow in the off-diagonal terms. This transformation is given by:
\begin{align}
U_{v_\mathrm{fr}}=\left[\begin{array}{c c}\exp\left(-i\frac{m}{\hbar}v_\mathrm{fr} x\right) &0\\0&\exp\left(i\frac{m}{\hbar}v_\mathrm{fr} x\right)\end{array}\right],
\end{align}
and using it leads to:
\begin{align}\label{eqn:Hfin}
 \tilde{H}^{(\pt)}_\BdG =\left[\begin{array}{c c}\hat{h}_{v_\mathrm{fr}}+\frac{\pt^2}{2m} & \Delta(x)e^{-i\frac{2m}{\hbar}v_{\bf}x}\\ \Delta^{*}(x)e^{+i\frac{2m}{\hbar}v_{\bf}x}&-\hat{h}_{-v_\mathrm{fr}}+\frac{\pt^2}{2m}\end{array}\right].
\end{align}
Here $v_\mathrm{fr}$ is associated with the unitary transformation, $v_\bf = v_\mathrm{s} - v_\mathrm{fr}$ has the interpretation of a backflow velocity, and $\hat{h}_{v_\mathrm{fr}} = \frac{\hbar^2}{2m}(-i\partial_x+m v_\mathrm{fr}/\hbar)^2 -\mu$. In this way the soliton velocity $v_\mathrm{s}$ is split into the velocity of the condensate backflow $v_\mathrm{bf}$ and $v_\mathrm{fr}$, which is a free parameter to be set in the calculations. Enforcing periodic (anti-periodic) boundary conditions in the $x$ direction, will ensure that $\arg[\Delta(L_x/2) / \Delta(-L_x/2)] - 2m v_\bf L_x/\hbar = 0$ ($\pi$) \footnote{Due to $\Delta(x)$ being a sum of products of the eigenstates of the BdG matrix, phase twists applied to either diagonal sector of BdG matrix ($\phi_u$ and $\phi_v$, respectively) add in contributing to $\Delta(x)$, while the individual phases do not have physical significance. In our calculations we usually set $\phi_u$ to $0$  and $\phi_v$ to $\pi$.}. 
In sufficiently large systems with $\pi$-twisted boundary conditions (as used through most of our calculations) $v_\mathrm{fr}$ and $v_\s$ differ only sightly, as $v_{{\bf}}\ll v_\mathrm{fr}\approx v_\s$.

Having found a self-consistent $\Delta(x)$ and thus knowing an approximate eigenfunction of the many-body Hamiltonian $\mathcal{H}$ one can calculate various observables of the system.  
The free energy  of the soliton-bearing system in the frame co-moving with  the soliton  is given by
\begin{align}
 F^{\mathrm{SF}}_\s&=\langle\mathcal{H}\rangle_\s 
 =\int\mathrm{d}x\sum_{\nu,\pt}' \varepsilon_{\nu}^{(\pt)} |u_{\nu}^{(\pt)}(x)|^2 ,
\end{align}
where the subscript s denotes system with a soliton, and the superscript SF means the soliton frame of reference.
In order to characterize the properties of the soliton itself it is more interesting to consider the change in free energy caused by adding a soliton rather than the (extensive) quantity $F^{\mathrm{SF}}_\s$, which is the free energy of the whole system.
Moreover, to compare different cases, the free energy has to be brought back to the laboratory frame of reference \cite{Shamailov2019a}.  Thus we have to calculate the difference between the free energy of system with ($F_\s$) and without ($F_0$) a soliton in the lab frame, and make an appropriate transformation to obtain a relation to the observables in the soliton frame (see App.~\ref{app:gb}). The soliton (free) energy is then defined as
\begin{align}
 \Delta F = F_\s-F_0=F_\s^{\mathrm{SF}}-F_0^{\mathrm{SF}}+N_\d m v_\s^2.
\end{align}
Here, $N_\d<0$ is particle number depleted by the soliton
\begin{align}
  N_\mathrm{d} = \int[n_\mathrm{s}(x)-n_0]\mathrm{d}\mathbf{r} =N_\s-N_0,
\end{align}
where $N_\s$ is the number of particles in the soliton-bearing system with density $n_\mathrm{s}(x)$, and $N_0$ is the number of particles in the system without the soliton with the constant background density $n_0$.

Next we can calculate the physical momentum of the system:
\begin{align}
  P^{\mathrm{SF}}_\mathrm{ph}&=
  \sum_\sigma\int\mathrm{d}\mathbf{r} \left\langle\psi\left|
  \psi_\sigma^\dag(\mathbf{r}) (-i \hbar \partial_x + mv_\mathrm{fr}) \psi_\sigma(\mathbf{r})
  \right|\psi\right\rangle 
   \nonumber\\
  &= 4\hbar\int\mathrm{d}x\sum_{\nu,\pt}'{v_{\nu}^{(\pt)*}}(x)\partial_x v_{\nu}^{(\pt)}(x).
\end{align}
In this frame of reference this is the sum of the physical momentum of the soliton and a momentum contribution from the $v_\mathrm{fr}$. Because we are using finite system sizes with an arbitrarily set phase twist $\phi_\t$ at the boundary, we have to add a boundary contribution  
to obtain the canonical momentum of the soliton
\begin{align}
P^{\mathrm{SF}}_\mathrm{c}=P^{\mathrm{SF}}_\mathrm{ph}+P^{\mathrm{SF}}_{\phi_\t},
\end{align}
where $P^{\mathrm{SF}}_{\phi_\t}=-\frac{\hbar}{2}\frac{\phi_\t}{L_x}N_\s$.
After transforming the momentum to the lab frame we obtain a more natural splitting:
\begin{align}
    P_{c} = P_\mathrm{ph}+P_{\phi_\s},
\end{align}
as $P_\mathrm{ph}$ is just physical momentum of the soliton: $P_\mathrm{ph}=N_\d m v_\s$, while $P_{\phi_\s}=-\frac{\hbar}{2}\frac{\phi_\s}{L_x}N_\s$ is proportional to the phase step of the soliton.

The dispersion relation of $\Delta F$ vs.~$P_\mathrm{c}$ of the soliton solutions is parameterized by the soliton velocity $v_\s$. Besides the interaction strength, these quantities also depend on the chemical potential $\mu$, which parameterizes the background particle number density. In the presence of a slowly-varying external potential (on the length scale of the soliton), the dispersion relations can be used to predict the motion of the soliton by treating the soliton like a quasiparticle \cite{Konotop2004}.
The free energy of the soliton at position $z(t)$ then depends on its velocity $v_\s = \dot{z}$ and the chemical potential at local equilibrium $\mu(z)$, which is modified by the external potential according to the local-density approximation. This leads to an equation of motion that resembles Newton's equation \cite{Konotop2004,Mateo2015}:

\begin{align} \label{eq:eom}
 \frac{1}{v_\s}\left.\frac{\partial \Delta F}{\partial v_\s}\right|_{\mu}\ddot{z_\s}=-m\left.\frac{\partial \Delta F}{\partial\mu}\right|_{v_\s}\frac{\partial\mu}{\partial z}.
\end{align}
The coefficients on the left and right sides of the equation can be interpreted as the inertial and the physical mass, respectively:
\begin{align}
M_\mathrm{P}&=m\left.\frac{\partial \Delta F}{\partial \mu}\right|_{v_\s},\label{eq:mp}\\
M_\mathrm{I}&=\frac{1}{v_\s}\left.\frac{\partial \Delta F}{\partial v_\s}\right|_{\mu}\label{eq:mi}.
\end{align}
Near the minima or maxima of the local chemical potential $\mu(z)$, the masses can be approximated as constants and Eq.~\eqref{eq:eom} describes harmonic oscillations. The frequency of these small-amplitude oscillations is then related to the mass ratio by $\omega_z^2/\Omega^2 = M_\mathrm{I}/M_\mathrm{P}$, where $\Omega$ is the harmonic oscillator frequency related to the extremum of the external potential \cite{Mateo2015}.

\section{Self consistent calculations\label{sec:scresults}}
The calculations were conducted for $N$ fermions confined in a two-dimensional box with dimensions $L_x\times L_y$ and periodic  (torus) boundary conditions
for a wide range of binding energies $E_\mathrm{b}$. To obtain a soliton solution we often apply an additional phase change of $\pi$ in the $x$ direction, realizing antiperiodic boundary condition, in order to offset the main contribution from the soliton phase step, as described in Sec.~\ref{sec:model}.

The renormalized interaction strength $g$ for a two-dimensional Fermi gas depends on $E_\mathrm{b}$ as
\begin{align}
 \frac{1}{g}=\sum_{p_{x}=-p_{x}^\mathrm{cutoff}}^{p_{x}^\mathrm{cutoff}}\sum_{p_{y}=-p_{y}^\mathrm{cutoff}}^{p_{y}^\mathrm{cutoff}} \frac{1}{\frac{1}{m} (p_{x}^2+p_{y}^2)+E_\mathrm{b}}.
\end{align}
where the values of $p_x$ and $p_y$ are integer multiples of ${2\pi\hbar}/{L_{x}}$ and ${2\pi\hbar}/{L_{y}}$, and $p_{x}^\mathrm{cutoff}$ and $p_{y}^\mathrm{cutoff}$ are the respective momentum cutoffs.
By changing $E_\mathrm{b}$ from $0$ to $+\infty$ we can obtain systems in both BCS and BEC regimes.
Additional calculations in one-dimensional systems were made for comparison. 

While presenting results we are using Fermi units based on the Fermi energy $\ef=\pi\hbar^2 N/mL_{x} L_{y} = \hbar^2 \kf^2/2m$, from which one can further derive the length scale $\lf = 2\pi/\kf$ and the velocity $\vf = \kf/m$. 
Other important quantities used throughout the paper are BCS correlation length $\xi_\mathrm{BCS}=\hbar \vf/\Delta_0$ and the pair-breaking velocity $v_\mathrm{pb}\equiv \Delta_0/\hbar\kf = \vf\sqrt{E_\mathrm{b}/2\ef}$ for the homogeneous two-dimensional Fermi superfluid with $\Delta_0=\sqrt{2E_\mathrm{b}\ef}$ and $\mu_0=\ef-E_\mathrm{b}/2$ \cite{Randeria1989}.

In the two dimensional system the transverse direction $y$ is assumed to be homogeneous. For the simulations we have used a square lattice discretization. Lattice spacings were set to $dL_{x}=dL_{y}=0.5\lf$ for $E_\mathrm{b}<\ef$.  For higher $E_\mathrm{b}$ and for few other cases when additional precision were needed the lattice spacings were set to $0.25\lf$. The lattice sizes were set to at least $L_{x}=15\xi_{BCS}$ for the inhomogeneous direction and $L_{y}=10\xi_{BCS}$ for homogeneous one in the BCS regime. In the BEC regime we used at least $L_{x}=40\lf$ and $L_{y}=30\lf$, respectively. Gradients were approximated using finite differences with a highly-accurate 9-point stencil.
Transverse momenta were summed in the range $\pt \in (-\pi\hbar/dL_{y},\pi \hbar/dL_{y}]$, with discretization $d\pt=2\pi \hbar/L_{y}$.

In order to obtain a self-consistent solution one has to start from some assumed form of the order parameter $\Delta_0(x)$ in order to construct an initial BdG matrix \eqref{eqn:Hfin}. The respective eigenstates are used to calculate the new order parameter using \eqref{eq:deltaPT} -- which we denote $\Delta_0'(x)$. In general $\Delta_0(x)\neq\Delta_0^\prime(x)$ and thus a new iterate $\Delta_1(x)$ has to be constructed leading to an iterative procedure. Convergence is achieved when $||\Delta_n -\Delta'_n||<\epsilon$ with some prescribed accuracy $\epsilon$. In our case the simplest self-consistent iteration scheme with $\Delta_n(x)=\Delta'_{n-1}(x)$ does not work, as it is unstable when applied to moving solitonic solutions (i.e.~when $v_\s\neq 0$). Instead, we use Broyden's method \cite{Broyden1965}, which is a quasi-Newton method where the Jacobian is not explicitly calculated at every step. As a starting point for the search we have used a step function or hyperbolic tangent (with the length scale $\xi_{\mathrm{BCS}}$). The initial Jacobians for Broyden's method were calculated explicitly using finite differences with a 2-point stencil. Broyden iterations were continued until the difference between the 2-norms and $\infty$-norms of $\Delta(x)$ and $\Delta'(x)$ were smaller than the threshold values, typically set-up around $\epsilon = 10^{-6}\ef$, which provided well converged results in most cases. All calculations were done using the Julia programming language \cite{Bezanson2017julia} and its standard linear algebra solvers based on LAPACK \cite{lapack99} were used for matrix diagonalization.

\begin{figure}
  \begin{center}
   \includegraphics[width=8cm]{"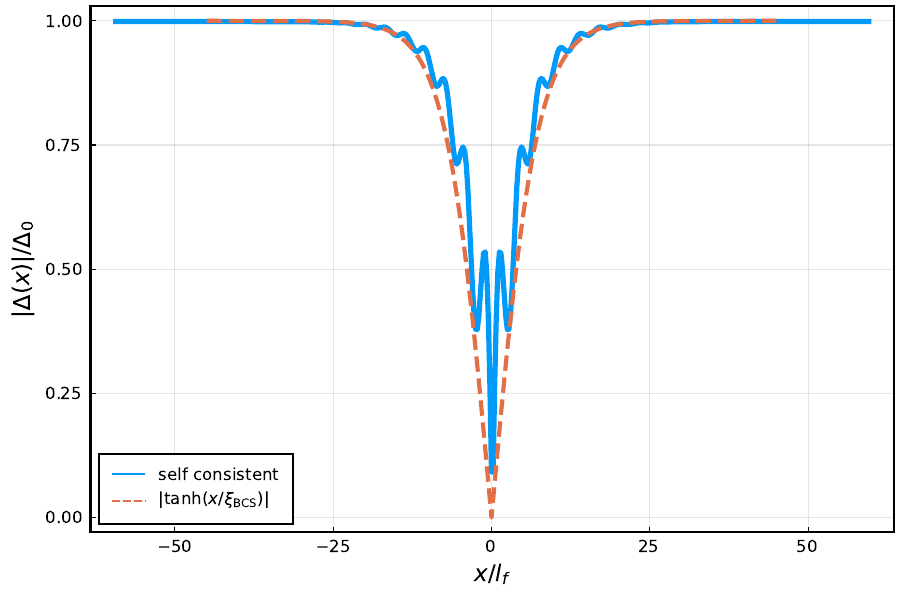"}
  \caption{Absolute value of the order parameter $|\Delta(x)|$ in a system with stationary dark soliton solution in the BCS regime of a two-dimensional Fermi superfluid.  The self consistent result is shown for $E_\mathrm{b}=0.02\ef$, $v_s\approx 0$ and $\phi_\mathrm{s}\approx \pi$ (blue solid line). The red dashed line is $|\tanh(x/\xi_{\mathrm{BCS}})|$.\label{fig:deltas}}
  \end{center}
 \end{figure}
 
A typical example of a self-consistent solution in the BCS regime with a soliton can be seen in  Fig.~\ref{fig:deltas}. For the absolute value $|\Delta(x)|$  we can observe a wide envelope given approximately by $|\tanh(x/\xi_{\mathrm{BCS}})|$ with additional oscillations  at the Fermi length scale $\lf$ and a central through, which is almost unaffected by the binding energy. We attribute the oscillations to Friedel oscillations and note that they appear also in quasi-particle eigenstates (not shown).
The general shape of the soliton order parameter is reminiscent of the dark soliton in the 3D Fermi superfluid at unitarity \cite{Antezza2007}. 

 \begin{figure}
  \begin{center}
   \includegraphics[width=8cm]{"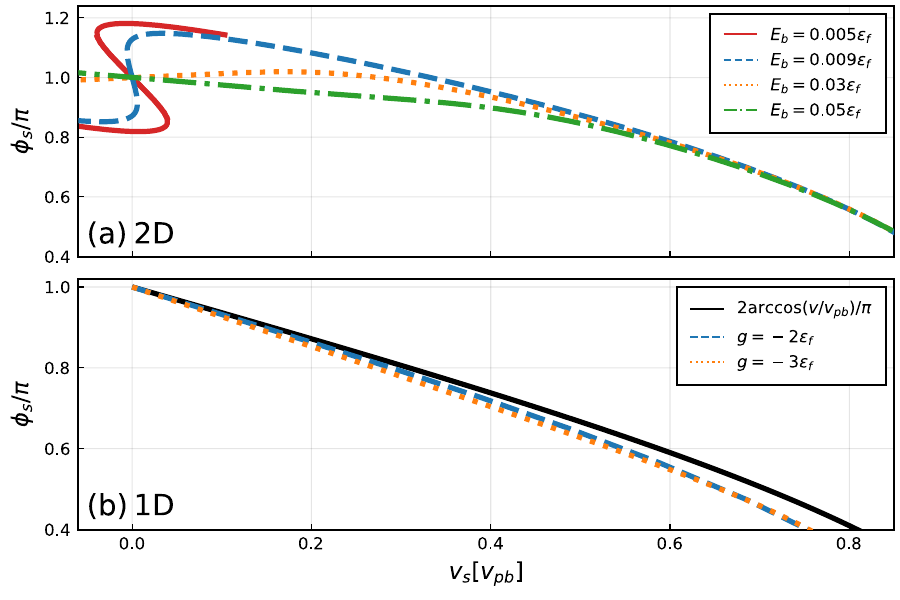"}
   \caption{The phase jump for families of dark/grey soliton solutions. Panel (a): Phase step of the soliton in the two-dimensional superfluid as a function of the soliton velocity for various binding energies: $E_\mathrm{b}=0.005\ef$ -- solid red, $E_\mathrm{b}=0.009\ef$ -- blue dashed, $E_\mathrm{b}=0.03\ef$ -- yellow dotted, $E_\mathrm{b}=0.05\ef$ -- green dash-dotted. Panel (b): The same relation for a one-dimensional BdG model (setting $\pt=0$) with $g=-2\ef$ -- blue dashed and $g=-3\ef$ -- yellow dotted lines. The black solid line is a theoretical prediction for a 1D model \cite{Efimkin2015}.\label{fig:phases}
   }
  \end{center}
 \end{figure}

In Fig.~\ref{fig:phases} we present phase-steps of the solitons as a  function of the soliton velocity $v_\s$ (scaled with pair-breaking velocity $v_\mathrm{pb}$) for various interaction strengths.
The soliton phase step is calculated as a difference between phases on the edges of the system transformed to a frame in which there is no backflow:
\begin{align}
 \phi_\s&={\arg(\Delta(x\rightarrow-\infty)/\Delta(x\rightarrow+\infty))}|_{v_\mathrm{bf}=0}\nonumber\\
 &\approx \left.{\arg\left(\Delta\left(-{L_x}/{2}\right)/\Delta\left({L_x}/{2}\right)\right)}\right|_{v_\mathrm{bf}=0}.\label{eq:phidef}
\end{align}

The phase step of a soliton is a characteristic feature of solitons in superfluids and is expected to depend on the soliton velocity. The top panel of Fig.~\ref{fig:phases} shows results for two-dimensional systems for several binding energies. Interestingly, for low binding energies we observe a dynamic change of the shape of the relation $\phi_\s(v_\s)$, which eventually leads to the appearance of three distinct solitonic solutions for the same velocity. It is also remarkable that for sufficiently high velocities all phase-step values seem to lie on one universal curve.

The bottom panel of Fig.~\ref{fig:phases} shows analogous results for one-dimensional systems, obtained by restricting our calculation to a single transverse momentum $p_\mathrm{t}=0$.
The numerical results are compared with the theoretical prediction from Ref.~\cite{Efimkin2015}, $\phi_\s=2\arccos(v_\s/v_\mathrm{pb})$ and show a qualitative agreement with this result. Note that Ref.~\cite{Efimkin2015} applied additional approximations to the BdG equations like a linearization of the fermion dispersion relation, which are not used in our calculations.

The differences between the situation in two and one dimensions as shown in Fig.~\ref{fig:phases} are striking. A further analysis of the change in the shape of the $\phi_\s(v_\s)$ relation will be given in the next section.

\section{Dispersion relations and inertial masses\label{soc:masses}}
Further we focus on the soliton masses and their dependence on the binding energy and velocity. The physical mass defined by Eq.~\eqref{eq:mp} can be also obtained from the depletion number \cite{Mateo2015}:
\begin{align}
 M_\mathrm{P} = N_\d m.  
\end{align}
The inertial mass can be calculated using Eq.~\eqref{eq:mi}. However, this equation  introduces significant numerical errors for $v_\s\approx0$. To alleviate this we are using the equivalent relation $M_\mathrm{I}=\sfrac{\partial P_\mathrm{c}}{\partial v_\s}$
(derived using: $v_\s =\sfrac{\partial \Delta F}{\partial P_\mathrm{c}}$). This equation can be further transformed into
\begin{align}
 M_\mathrm{I}=\frac{\partial P_\mathrm{c}}{\partial v_\s} =\frac{\partial P_\mathrm{ph}}{\partial v_\s}+\frac{\partial P_{\phi}}{\partial v_\s}
 =M_\mathrm{P} + mL_{y} \frac{\partial \phi_\s}{\partial v_\s}.
\end{align}
It shows that the difference between the inertial and the physical mass is proportional to the slope of the $\phi_\s(v_\s)$ relation presented in Fig.~\ref{fig:phases}.
    
\begin{figure}
 \begin{center}
  \includegraphics[width=8cm]{"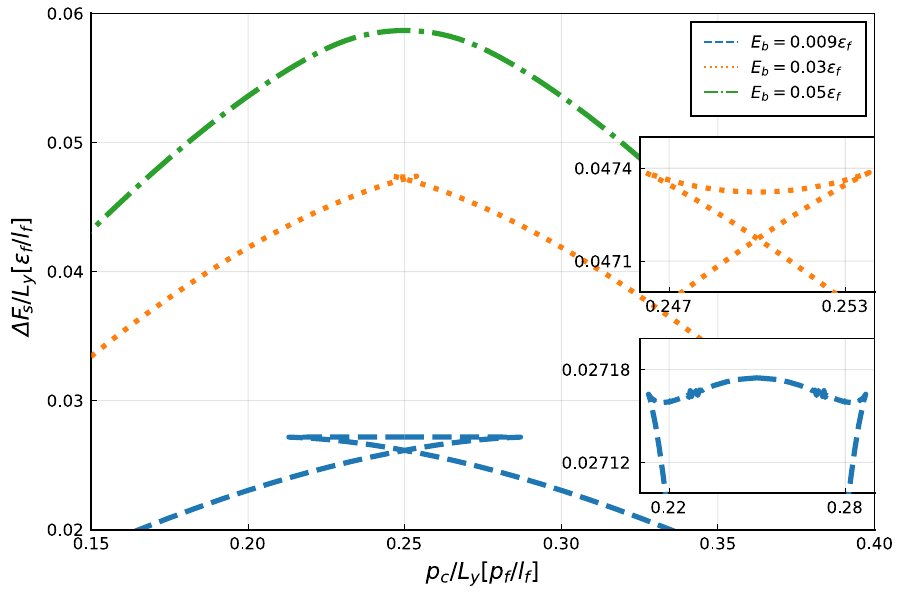"}
  \caption{Soliton dispersion relations. Free energy of the soliton  as a function of canonical momentum (scaled with transverse system length). Results for $E_\mathrm{b}=0.009\ef, 0.03\ef, 0.05\ef$ are shown with the dashed blue, dotted yellow and dash-dotted green lines, respectively. Insets present closeups on the features of the $E_\mathrm{b}=0.03\ef$ and $E_\mathrm{b}=0.009\ef$ curves.\label{fig:dispersion}}
 \end{center}
\end{figure}

The soliton energy -- momentum dispersion relations are presented for a range of binding energies in Fig.~\ref{fig:dispersion}, using the same color scheme as in Fig.~\ref{fig:phases}. The quantities $\Delta F_\s$, $P_\mathrm{c}$, $M_\mathrm{I}$ and $M_\mathrm{P}$ are extensive in $y$ due to the homogeneity of the system in this direction. Thus we are plotting the scaled,  intensive values $\Delta F_\s/L_{y}$, $P_\mathrm{c}/L_{y}$, $M_\mathrm{I}/L_{y}$ and $M_\mathrm{P}/L_{y}$.

As one can see in Fig.~\ref{fig:dispersion}, there are qualitative differences between the presented dispersion curves. For $E_\mathrm{b} = 0.05\ef$ we have a smooth inverted lobe with negative curvature. For $E_\mathrm{b} = 0.03\ef$ the inverted lobe is split in the center by two cusps. The outer arms retains a negative curvature but the central part is convex.
For even smaller $E_\mathrm{b}=0.009\ef$, the curvature of the central part changes. In the very center the curvature is negative while it is positive at the sides. There are two inflection points where the curvature vanishes and the inertial mass diverges according to \eqref{eq:mi}.

\begin{figure}
 \includegraphics[width=8cm]{"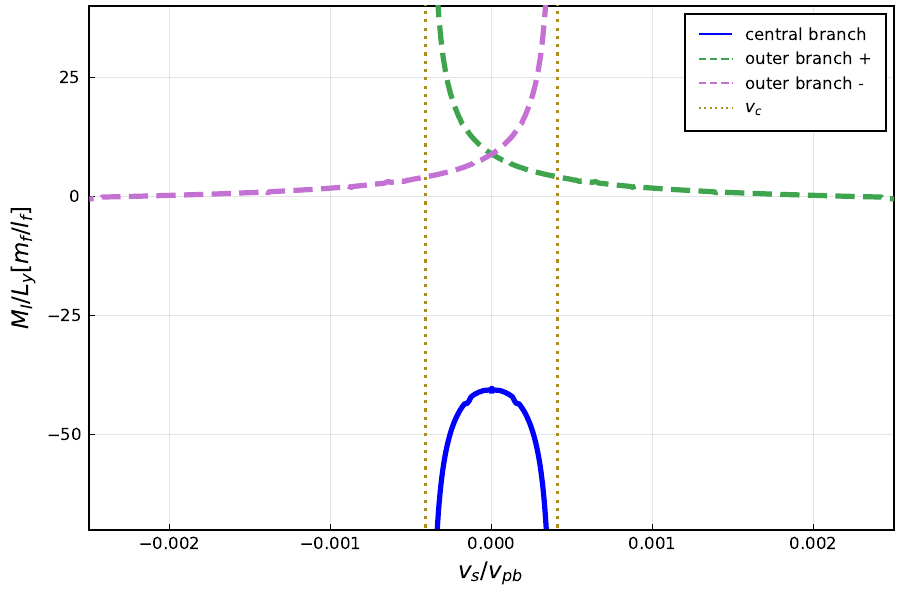"}
 \caption{Inertial mass for $E_\mathrm{b}=0.009\ef$ as function of the soliton velocity. The blue solid line is the central branch, which contains the $\pi$-phase stationary soliton. The green and purple dashed lines are the outer branches containing non-$\pi$-phase stationary solutions. Yellow vertical dotted lines denotes the critical velocity $v_\mathrm{c}$ at which the inertial mass diverges.\label{fig:MI0009}}
\end{figure}

To better understand the latter, most complicated, case we plot the inertial mass as a function of the soliton velocity in Fig.~\ref{fig:MI0009}. The central branch (depicted with solid blue line) is symmetric around $v_\s=0$. It always stays negative and diverges to $-\infty$ for $|v_\s|\rightarrow v_\mathrm{c}$, where $v_\mathrm{c}$ is the critical velocity at which the curvature of the dispersion relation vanishes. 

One of the outer branches (green line) starts at $-v_\mathrm{c}$ where it has $+\infty$ singularity, then it monotonically decreases, eventually crossing $0$ (which is not well visible on the plot). The other is a mirror image of the first with opposite velocities.

\begin{figure}
 \begin{center}
  \includegraphics[width=8cm]{"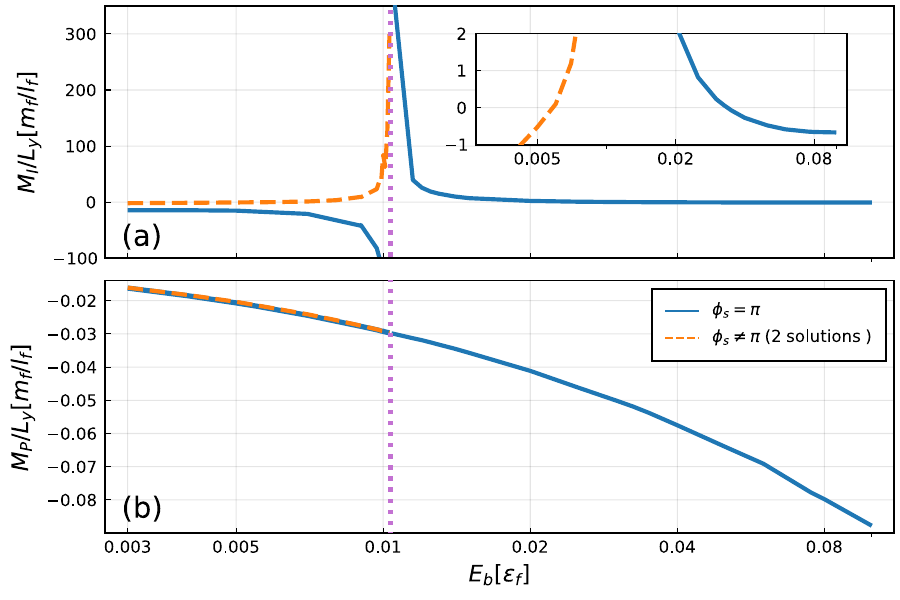"}
  \caption{Soliton mass as function of the pair binding energy. (a) Inertial and (b) physical mass per transverse length for stationary ($v_\s=0$) solitons for a range of binding energies. The blue solid line shows data for the soliton solution with $\pi$-phase step, while the orange dashed line shows the mass for the two degenerate soliton solutions with non-$\pi$ phase step. The dotted magenta line marks the value of divergence $E_\mathrm{b}^{\infty}=0.01025\pm0.0005 \ef$.
  The inset presents a closeup of the two upper curves to show that they both are crossing $M_\mathrm{I}=0$ at some point.\label{fig:masses}}
 \end{center} 
\end{figure}

As there is a particular interest in slowly moving solitons, we consider the stationary ($v_\s = 0$) solutions specifically, and plot their masses  for a range of binding energies in Fig.~\ref{fig:masses}. The inertial mass is shown in panel Fig.~\ref{fig:masses}(a) and the physical mass in panel Fig.~\ref{fig:masses}(b).

Analyzing the data presented in Fig.~\ref{fig:masses} we can mark three transition points: At $E_\mathrm{b}^{0}=0.034\pm0.0005\ef$ the inertial mass vanishes for the sole $v_\s=0$ soliton solution (with $\phi_\s=\pi$).  At the smaller binding energy $E_\mathrm{b}^{\infty}=0.01025\pm0.0005 \ef$ signifying weaker interactions, the effective mass of all $v_\s=0$ soliton solutions diverge. Finally at even weaker interactions with  $E_\mathrm{b}^{0\prime}=0.006\pm0.001\ef$ a third transition point is found where the inertial mass vanishes for two $v_s=0$ solutions with non-$\pi$ phase step.

Looking from the right-hand side, for high binding energies $E_\mathrm{b}>E_\mathrm{b}^{0}$ the inertial mass is negative. Below  $E_\mathrm{b}^{0}$ it becomes positive and grows towards infinity approaching $E_\mathrm{b}\rightarrow E_\mathrm{b}^{\infty (+)}$. On the other side of this singularity for $E_\mathrm{b}<E_\mathrm{b}^{\infty}$ we have three distinct solutions with stationary solitons. One branch associated with $\pi$-phase solutions  diverges to $-\infty$ for $E_\mathrm{b}^{\infty (-)}$, while for lower $E_\mathrm{b}$ values it is finite and negative. The other two degenerate curves are associated with non-$\pi$-phase solutions. These solutions only exist for $E_\mathrm{b} < E_\mathrm{b}^{\infty}$ and their inertial mass diverges to $+\infty$ at the termination point $E_\mathrm{b}^{\infty}$. 
Further decreasing $E_\mathrm{b}$, the mass of the non-$\pi$-phase solitons decreases monotonically eventually becoming negative for $E_\mathrm{b} < E_\mathrm{b}^{0\prime}$.

The physical mass shown on the lower panel does not exhibit as dramatic changes as the inertial mass. For all binding energies in our numerical data it remains negative and decreases monotonically with increasing $E_\mathrm{b}$. On the BEC side of the crossover it decreases, while towards the BCS regime it increases towards zero, which is consistent with the observation that we do not observe any change in the particle number density caused by the presence of the soliton in that limit. No noticeable abrupt change of the physical mass can be seen upon crossing the threshold $E_\mathrm{b}^{0}$. Below $E_\mathrm{b}^{\infty}$ three different branches appear but differences in $M_\mathrm{P}$ between the $\pi$-phase solution and the others are orders of magnitude smaller than the actual $M_\mathrm{P}$ values.

\begin{figure}
 \begin{center}
  \includegraphics[width=8cm]{"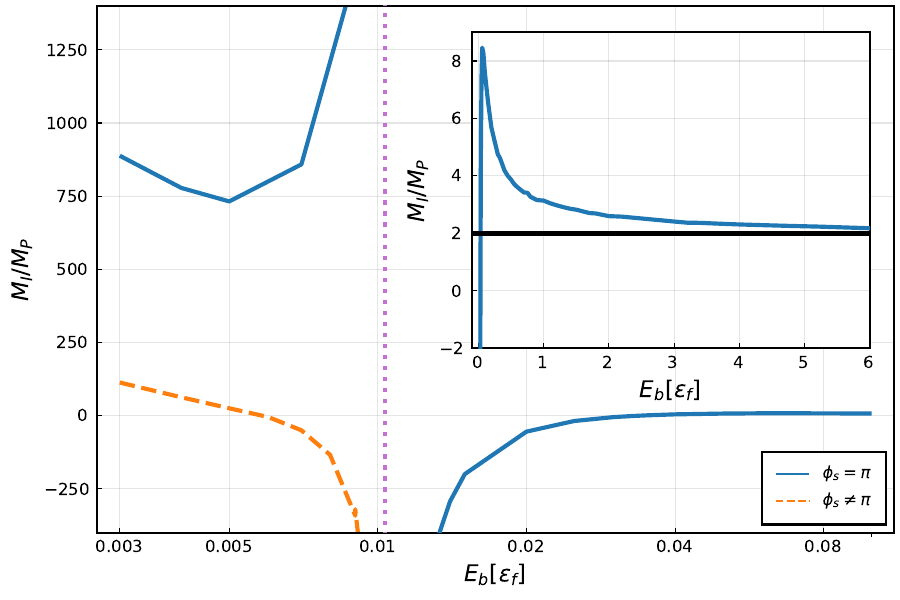"}
  \caption{Ratio of inertial and physical mass for stationary ($v_\s=0$) solitons for a range of binding energies. The inset shows the only branch for $E_\mathrm{b}>E_\mathrm{b}^{0}$ for a wider range of binding energies. The black horizontal line is the known value for BEC limit $M^{\mathrm{BEC}}_\mathrm{I}/M^{\mathrm{BEC}}_\mathrm{P}=2$ \cite{Busch2000,Konotop2004}. 
  Otherwise the line styles and color schemes are as in Fig.~\ref{fig:masses}.
  \label{fig:ratio}}
 \end{center} 
\end{figure}

In Fig.~\ref{fig:ratio} we present the ratio of the inertial and the physical mass as a function of the binding energy. Due to the qualitatively simple dependence of $M_\mathrm{P}$ on $E_\mathrm{b}$, the behavior of the ratio $M_\mathrm{I}/M_\mathrm{P}$ is governed mostly by $M_\mathrm{I}(v_\s)$. 
An interesting takeaway is that the ratio appears to correctly approach the value of 2 for large $E_\mathrm{b}$ in agreement with predictions from Gross-Pitaevskii theory for a BEC of strongly-bound bosons \cite{Busch2000,Konotop2004}, while we know that BdG mean-field theory in two dimensions fails to correctly predict the equation of state in the BEC regime \cite{Bertaina2011}.

Before summarizing the three different physical regimes, we list a few general rules that are useful when interpreting the presented data:
\begin{itemize}
 \item The value of the inertial mass $M_\mathrm{I}$ is largely determined by the slope of $P_\mathrm{c}(v_\s)\propto-\phi_\s(v_\s)$ and can be alternatively obtained from the inverse curvature of the dispersion relation $\Delta F_\s(P_\mathrm{c})$.
 \item The soliton velocity $v_\s$ is determined by the slope of the dispersion relation through $v_\s = d\Delta F_\s(P_\mathrm{c})/d P_\mathrm{c}$.
 \item Therefore, smooth extrema of the dispersion relation mark points where $v_\s=0$, i.e.\ stationary soliton solutions exist.
 \item When the dispersion relation has a cusp the slope of the relation $P_\mathrm{c}(v_\s)$ vanishes and consequently $M_\mathrm{I}=0$.
 \item When the dispersion relation has an inflection point the curve $P_\mathrm{c}(v_\s)$ has a vertical asymptote and the inertial mass $M_\mathrm{I}$ diverges.
\end{itemize}

\emph{Tightly-bound fermion pairs:}
For $E_\mathrm{b}>E_\mathrm{b}^{0}$ (green dash-dotted line in Figs~\ref{fig:phases} and \ref{fig:dispersion}) the canonical momentum is monotonous as a function of the soliton velocity ($v_\s$). The dispersion relation is an inverted lobe and the inertial mass is always negative. Qualitatively, the behavior of systems in this regime resembles the one in the BEC limit. 

When $E_\mathrm{b}=E_\mathrm{b}^{0}$, the relation $P_\mathrm{c}(v_\s)$ flattens for $v_\s=0$ and a cusp appears in the center of the dispersion relation, which means that $M_\mathrm{I}(v_\s=0)=0$.

\emph{Simple swallow-tail regime with positive inertial mass:}
For $E_\mathrm{b}^{0}> E_\mathrm{b} >E_\mathrm{b}^{\infty}$ (yellow dotted line in Figs~\ref{fig:phases} and \ref{fig:dispersion}), the dispersion relation has a swallow-tail shape. It is split into three branches by two cusps. The outer branches are symmetric and have negative curvature while the central one has positive curvature. The slope at the cusp defines a threshold velocity $v_\t$ at which the sign of $M_\mathrm{I}$ changes. The canonical momentum decreases with growing $v_\s$ in the central part (corresponding to the central branch of the dispersion relation) and increases for $|v_\s|>v_\t$. The inertial mass is positive for the central branch and negative for the rest.

At the point $E_\mathrm{b}=E_\mathrm{b}^{\infty}$ the dispersion relation retains a swallow-tail shape but its curvature drops to $0$ at $v_\s=0$. The $P_\mathrm{c}(v_\s)$ relation acquires a vertical asymptote. The inertial mass is positive  for $|v_\s|<v_\t$ and grows to indefinitely as $v_\s$ approaches $0$.

\emph{Complex swallow-tail regime:}
For $E_\mathrm{b}<E_\mathrm{b}^{\infty}$ (blue dashed line in Figs~\ref{fig:phases} and \ref{fig:dispersion}), the central branch in the dispersion relation has a maximum in the center, indicating a stationary soliton solution with negative inertial mass. Two other symmetric extrema exist. Depending on the value of $E_\mathrm{b}$ they appear in the central or the outer branches and are minima or maxima respectively. The $P_\mathrm{c}(v_\s)$ relation becomes three-valued in the central part. This leads to the appearance of a range of velocities around $v_\s=0$ for which there are three distinct solutions.  Only one $v_\s=0$ solution has $\phi_\s=\pi$ while the other two have phase steps that are symmetric around $\pi$. This leads also to existence of three separated branches of $M_\mathrm{I}(v_\s)$ as seen in Fig.~\ref{fig:MI0009}.

The rich scenario for the inertial mass with zero crossings and divergences in the same dispersion relation as seen in Fig.~\ref{fig:MI0009} has interesting consequences for the soliton dynamics in a trapped Fermi gas. It is important to note, though, that the Newton-like equations for quasiparticle-like soliton dynamics \eqref{eq:eom} with the mass parameters of Eqs.~\eqref{eq:mp} and \eqref{eq:mi} only hold for a slowly varying potential and assumes the adiabatic adjustment of the soliton to its environment. In the case of rapid changes of the inertial mass under small changes of velocity these conditions may be compromised and the quasiparticle picture of the soliton dynamics may break down.

An important physical consequence of the sign of the effective mass is that it predicts the presence (or absence) of the snaking instability. The snaking instability is a dynamical instability of a soliton solution in two or more dimensions, where an initially homogeneous solution (a soliton stripe) in the transverse direction spontaneously disintegrates through a snaking process \cite{Kuznetsov1988,Muryshev1999,Brand2002,cetoliSnakeInstabilityDark2013}.
The snaking instability can occur when the inertial mass $M_\mathrm{I}$ is negative, which leads to unstable small-amplitude oscillations of sinusoidal modulations of the soliton shape in the presence of a restoring force provided by the surface tension coming from the energy density of the soliton $\Delta F_\s/L_y$ \cite{Kamchatnov2008}. The snaking instability is not present when the inertial mass is positive,  in which case stable small amplitude oscillations are obtained. Thus, the regions of negative inertial mass in  Fig.~\ref{fig:MI0009} indicate unstable solutions under the snaking instability, while positive $M_\mathrm{I}$ solutions are stable.

\section{Counterflow in momentum space\label{sec:counterflow}}

The swallow-tail scenario for the dispersion relations of a soliton in a two-dimensional Fermi superfluid presents an unexpected behavior and has no analogy in bosonic superfluids. While swallow-tail dispersions are known to occur for dark solitons in a BEC in the presence of a periodic potential \cite{Mueller2002b,munozmateoNonlinearWavesBoseEinstein2019a}, or for transversely modulated vortex-ring solutions \cite{Komineas2003, Mateo2015}, they are not present for transversely homogeneous solutions in a flat potential \cite{Tsuzuki1971}. Previous studies of fermionic superfluids with the BdG equation in one \cite{Efimkin2015} or three dimensions \cite{Liao11pr:FermiSolitons,Scott2011,Spuntarelli2011} also have not shown any such effects. The numerical studies in three dimension have not explored the deep BCS regime, however, due to the numerical challenges posed by resolving the diverging length scales of the soliton envelope (the coherence length) and the Fermi length scale.

In order to properly understand what is happening it would be convenient to find analytic solutions of solitons in the BdG equations. Such solitons are available in a one-dimensional setting  \cite{Efimkin2015}, where, however, smooth dispersion relations with a very simple structure were predicted (as shown in Fig.~\ref{fig:phases}). The one-dimensional analytical solutions, which rely on linearized single-particle dispersions, cannot be simply extended to two spatial dimensions due to the contributions from transverse momenta.

To better understand source of this difference and its connection with higher dimensionality of our problem we will consider the case of $E_\mathrm{b}=0.005\ef$ where three stationary (i.e.\ $v_\s=0$) solutions exist with different phase steps.

\begin{figure}
\begin{center}
 \includegraphics[width=8cm]{"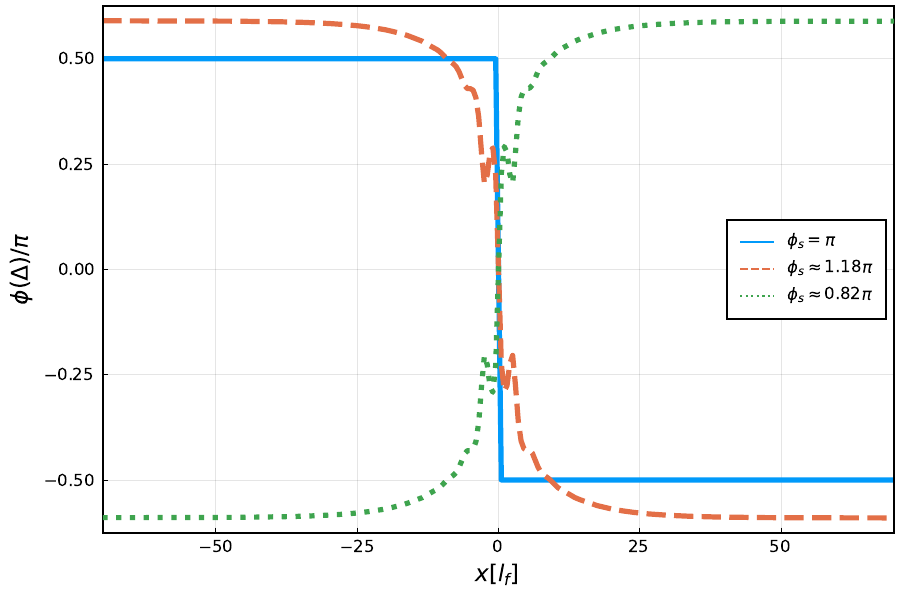"}
\caption{Complex phases of $\Delta(x)$ for solitonic self-consistent solutions. All three results are stationary and are obtained for the same value of $E_\mathrm{b}=0.005\ef$. The blue solid line is a soliton with $\phi_\s=\pi$. The red dashed and green dotted lines are results with non-$\pi$ phases.\label{fig:phases_deltas}}
 \end{center}
\end{figure}
In Fig.~\ref{fig:phases_deltas} we show the complex phase of $\Delta(x)$ of the solitonic solutions. Two qualitatively different behaviors can be observed. The $\pi$-phase solution simply has a constant phase with a discontinuous phase jump of exactly $\pi$. The solution can thus be presented as a purely real function. The other two solutions, however, are irreducibly complex with a continuous change of phase. One of these solutions is the complex conjugate of the other.

It is known that for bosonic superfluids the gradient of the phase is directly connected with the superfluid velocity, which is proportional to the current by the particle number density \cite{Landau1965}. The same connection can also be made for fermionic superfluids in case of a homogeneous flow. It is obviously not the case here, as we observe nonzero phase gradients for the stationary soliton solutions.
\begin{figure}
 \begin{center}
  \includegraphics[width=8cm]{"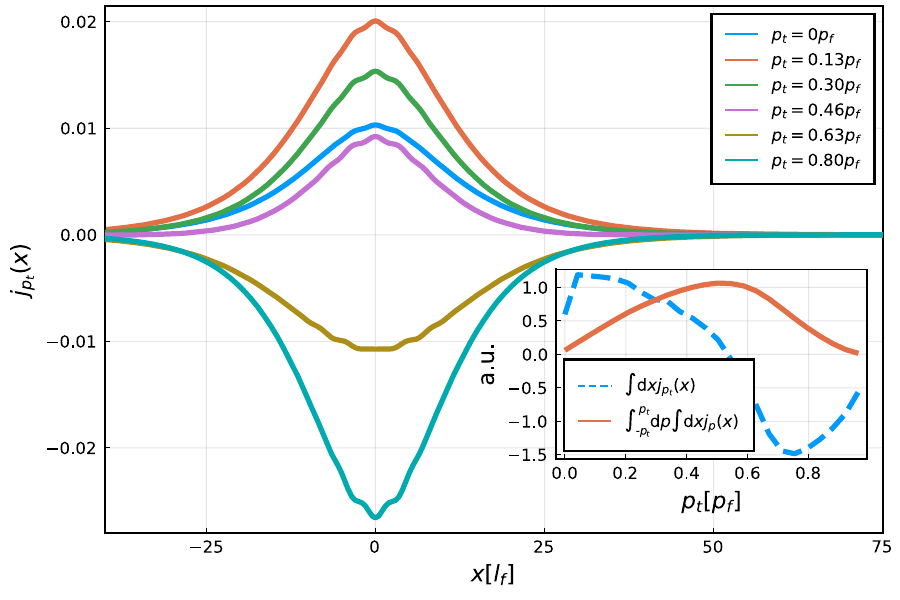"}
  \caption{Current densities for different transverse momenta for $E_\mathrm{b}=0.005\ef$, $v_\s=0$, $\phi_\s=0.82\pi$. In the inset current values integrated over $x$ are presented as function of $\pt$ are plotted as dashed blue line. The solid red line is a cumulative sum for momenta from $-\pt$ to $\pt$ (currents are symmetric in transverse momenta).\label{fig:caption2}}
 \end{center}
\end{figure}

In order to understand this discrepancy let us recall that the soliton solutions are inhomogeneous only in one dimension but a second, homogeneous, transverse direction is present as well. In the BdG formalism the transverse direction is included by summing multiple one-dimensional solutions with different transverse momenta $\pt$, as in Eq.~\eqref{eq:deltaPT} to obtain the complex order parameter $\Delta(x) = \sum_{\pt} \Delta_{\pt}(x)$. One can calculate the partial contributions $\Delta_{\pt}(x)$ as well as the transverse contributions  $j_{\pt}(x)$ to the current density, where $j(x) = \sum_{\pt} j_{\pt}(x)$ for each transverse momentum channel:
\begin{align}
\Delta_{\pt}(x)&=-g\sum_\nu^{\prime} u_\nu^{(\pt)}(x)v_\nu^{(\pt)*}(x),\\
j_{\pt}(x)&=-2\hbar\sum_{\nu}v_\nu^{(\pt)*}(x)\partial_x v_\nu^{(\pt)}(x).
\end{align}
Inspecting the different transverse contributions reveals that the partial current densities  $j_{\pt}(x)$ have different signs depending on $\pt$. Consequently, the current contributions have different  directions. This  can be seen in Fig.~\ref{fig:caption2}. In the special case we are considering now -- the stationary soliton -- all the flows add up to zero. This can be seen in the inset of Fig.~\ref{fig:caption2}. The blue dashed line represents the integrated current densities as a function of the transverse momentum.
One can see that it changes sign -- for small $|\pt|$ we observe flow in one direction while for larger $|\pt|$ the direction reverses. The red line represents a cumulative sum. It goes to zero for large transverse momentum, which demonstrates that eventually the total current vanishes as we should expect for a stationary soliton.

\begin{figure}
 \includegraphics[width=8cm]{"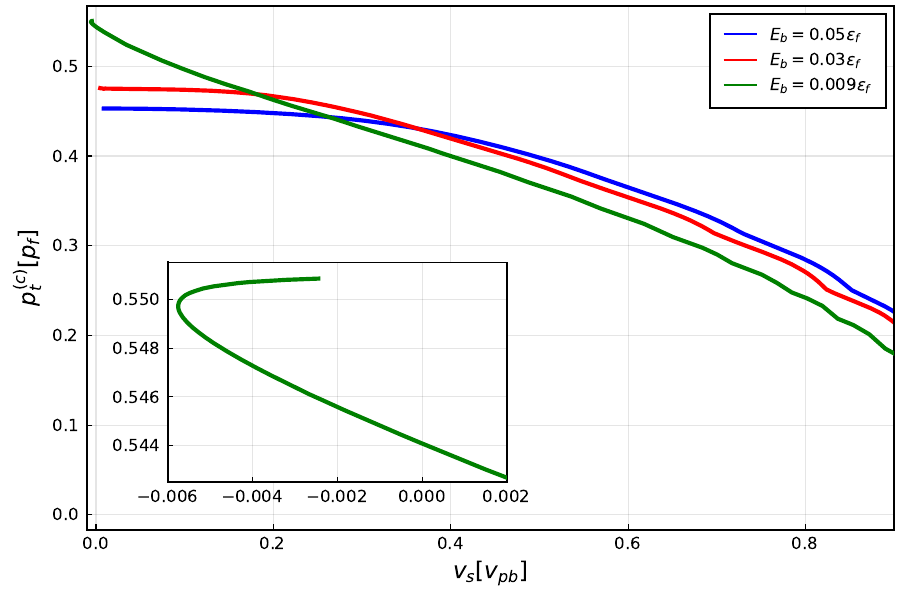"}
 \caption{Sign changing current contributions in the soliton solutions to the BdG equations. Shown is the 
 threshold transverse momentum $p^{(c)}_\t$ (in units of the Fermi momentum) above which the current contribution changes sign. It is presented as a function of the soliton velocity (scaled with the pair breaking velocity) for three different values of the binding energies. The inset shows a closeup for $E_\mathrm{b}=0.009\ef$ and for small $v_s$.\label{fig:thrshld}}
\end{figure}

The example presented above showed that stationary soliton solutions with 
finite phase gradients and a non-$\pi$ phase-step can exist and are supported by a 
cancellation of the partial flows for different transverse momenta. However, the coexistence of different partial flow directions is not limited to the  specific soliton solution that was presented. Such coexistence of flows with different directions is found also for moving solitons and for a much wider range of binding energies, up to $E_\mathrm{b}\approx0.2\ef$.
In Fig.~\ref{fig:thrshld} we show the approximate threshold of the transverse momentum $p^{(c)}_\t$ at which the transition between different flow directions occurs for a range of velocities and for three different binding energies.

\section{Conclusions\label{sec:conclusion}}
We have presented numerical simulations of solitons in a two-dimensional Fermi superfluid. The results are obtained as self-consistent solutions from the Bogoliubov-de Gennes method.
For the small binding energies on the far BCS side of the crossover regime we have observed unexpected and previously unseen behavior of the solitons.

Below a threshold binding energy of $E_\mathrm{b}^{0}=0.034\pm0.0005\ef$ the dispersion relations change shape from smooth lobes to swallow tail type, which also leads to a change of sign of the inertial mass of the soliton. Below the even lower threshold of $E_\mathrm{b}^{\infty}=0.01025\pm0.0005\ef$  we observe the  coexistence  of several solutions for the same velocity, as well as the existence of velocities for which the inertial mass diverges to infinity.
Those phenomena are connected with the fact that different one-dimensional solutions contributing to the full soliton shape contribute partial superfluid currents flowing in opposite directions. This allows, for example, a situation where a soliton with a nontrivial complex phase profile can be stationary, as flows for different transverse momenta cancel each other.

The soliton solutions studied in this work, if realized experimentally, can show a range of interesting properties. An example is the coexistence of multiple solutions with the same velocity that react differently to an external potential gradient. Another example are non-standard oscillations in a harmonic external potential due to a change of sign of the inertial mass, or a sudden breakdown of solitons when  they approach a velocity where the inertial mass vanishes.

In the model studied in this paper the transverse direction is assumed to be homogeneous. However, a similar setup could be obtained for ultracold atoms in a flat-bottom trap. This might allow for an experimental detection of the current components with different signs for different transverse momenta predicted by our numerical simulations. The different transverse momentum components can be separated in an experiment by time-of-flight position measurements after turning off the trap in the $y$ direction.
Previously overlapping components with different $p_\t$ will be separated. Then different directions of flow of the different components could be observed, which may be a method to experimentally confirm this phenomenon.

\section*{Acknowledgments}
The authors thank Uli Z\"ulicke for stimulating discussions.
J.M.\ wants to acknowledge support of National Science Centre (Poland)
via Sonatina grant 2020/36/C/ST2/00494. J.B.'s work was supported by the Marsden Fund of New Zealand, Contract No.\ MFP-MAU2007 from government funding managed by the Royal Society of New Zealand Te Apārangi.

\appendix
\section{Chemical potential of moving superfluid\label{app:mu}}
Let us consider a frame transformation where we move with velocity $v_\mathrm{fr}$ in the $x$ direction. 
It is important to note that we have to change the value of the chemical potential $\mu$, if we want to preserve the density in the moving frame of reference. The reason for this can be easily understood considering the simpler model of a homogeneous fermionic superfluid in a box with periodic boundary conditions carrying a ring-current. For given binding energy $E_\mathrm{b}$ (parameterizing the coupling strength) we can self consistently find $\Delta_0$. Constraining the density to be constant fixes the chemical potential $\mu$, which also has to be determined self consistently. For a stationary superfluid in two dimensions the values are given by $\Delta_0=\sqrt{2E_\mathrm{b}\ef}$ and $\mu_0=\ef-E_\mathrm{b}/2$ \cite{Randeria1989}.

A ring current can be introduced by adding a linear phase change to $\Delta(\r)=\Delta e^{i k_\mathrm{fr} x}$ (where $2m v_\mathrm{fr}/\hbar = k_\mathrm{fr}= 2\pi n/L_{x},\;n\in\mathbb{N}$, and $L_{x}$ is the length of the box in $x$ direction). After absorbing the phase gradient in equal portions in the $u$ and $v$ quasiparticle amplitudes, the BdG matrix of such a system can  be  represented in momentum space as
\begin{align}\label{eqn:Hkflow}
  H_k=\left[\begin{array}{c c}\xi_{{k}+k_\mathrm{fr}/2} & \Delta\\ \Delta^*&-\xi_{{k}-k_\mathrm{fr}/2}\end{array}\right],
\end{align}
where we have suppressed the transverse momentum indices (formally set $\pt=0$) for simplicity.
As the one particle energy reads $\xi_{{k}} = \frac{\hbar^2}{2m}k^2 -\mu$, we have
\begin{align}
\xi_{{k\pm k_\mathrm{fr}/2}} = \frac{\hbar^2}{2m}k^2 \pm \frac{\hbar^2}{2m}kk_\mathrm{fr} +\frac{\hbar^2}{2m}\frac{k_\mathrm{fr}^2}{4} -\mu.
\end{align}
The quadratic term can be grouped together with chemical potential
\begin{align}
  \tilde{\mu}=\mu-\frac{\hbar}{2m}\frac{{k_\mathrm{fr}}^2}{4},
\end{align}
while mixed term $kk_\mathrm{fr}$ has the same sign in both particle and hole segments. As an energy shift it affects only the quasiparticle energies but not eigenstates. As long as this energy shift is smaller than the half gap $\Delta/2$ it will not change quasiparticle level occupation numbers. Thus we have
\begin{align}\label{eq:current_eig1}
\varepsilon_{{k}}&=\pm\left(\frac{\hbar^2}{2m}{k}k_\mathrm{fr}+\sqrt{\tilde{\xi}_{{k}}^2+\Delta^2}\right),\\ \label{eq:current_eig2}
  u_{{k}+k_\mathrm{fr}/2}&=\frac{1}{\sqrt{2}}\left(1+\frac{\tilde{\xi}_{{k}}}{\sqrt{\tilde{\xi}_{{k}}^2+|\Delta|^2}}\right)^{1/2},\\
  \label{eq:current_eig3}
  v_{{k}-k_\mathrm{fr}/2}&=\frac{1}{\sqrt{2}}\left(1-\frac{\tilde{\xi}_{{k}}}{\sqrt{\tilde{\xi}_{{k}}^2+|\Delta|^2}}\right)^{1/2},
\end{align}
where
\begin{align}
  \tilde{\xi}_{{k}}=\frac{\hbar^2}{2m}{k}^2 -\tilde{\mu}.
\end{align}
Note that the particle number density is given by Eq.~\eqref{eq:n}.
We choose to keep it fixed and determine the chemical potential $\mu$ self-consistently.
As the equations \eqref{eq:current_eig2} and \eqref{eq:current_eig3} have the same form as equations without the current, a self-consistent solution will yield the same values:
\begin{align}
  \Delta&=\Delta_0,\\
  \tilde\mu&=\mu_0.
\end{align}
Going back to the original variables:
\begin{align}
\mu&=\tilde{\mu}+\frac{\hbar^2}{2m}{k_\mathrm{fr}}^2, \label{eq:muchange}
\end{align}
where $\mu$ is the chemical potential in the frame co-moving with the ring current, i.e.~where no explicit current is present. $\tilde{\mu}$ is the chemical potential in the frame where a current with the superfluid velocity $v_\mathrm{fr} = \hbar k_\mathrm{fr}/2m$ is present. 
So if we want to keep the same density in cases with different ring-currents (but also with backflow caused by the presence of soliton) we have to change chemical potential in accordance with Eq.~\eqref{eq:muchange}.
Another way to look at this phenomenon is to consider Fermi spheres for a moving superfluid. Due to a velocity-induced shift between particle and hole lobes, if the chemical potential remained constant, size of the Fermi sphere would be smaller, which would lead to a smaller density. To preserve the density, the Fermi sphere size has to be increased by adjusting the chemical potential.

\section{Observable transformation under Galilean boost\label{app:gb}}
In the main part of this article self-consistent solutions are obtained in the frame co-moving with the soliton. This stems from the construction of the time-independent method. The observables calculated using those results are also in the co-moving frame. 
However, to analyze and compare those results, this data have to be transformed back to the lab frame. For Galilean boost energy and momentum transform as follows:
\begin{align}
    P^\mathrm{SF}&=P - M v_\s, \\
    E^\mathrm{SF}&=E +\frac{1}{2}M v_\s^2 - P v_\s,
\end{align}
where we are using the same notation as in the main article: superscript $\mathrm(SF)$ for soliton frame and no superscript for lab frame. As we are working in the grand canonical ensemble instead of energy we are using free energy of the system:
\begin{align}
    F^\mathrm{SF}=E^\mathrm{SF}-\mu_{v_\s} N
\end{align}
While density is kept constant, the chemical potential will change as shown in App.~\ref{app:mu}:
\begin{align}
    \mu_{v_\s} = \mu_0 + \frac{1}{2}m v_\s^2,
\end{align}
if we assume that in the lab frame we do not have backflow. Further: 
\begin{align}   
    F^\mathrm{SF}+\mu_{v_\s} N &= F +\mu_0 N+\frac{1}{2}M v_\s^2 - P_c v_{v_\s},\nonumber\\
    F^\mathrm{SF} +\mu_0 N + \frac{1}{2}M v_{v_\s}^2&= F_0 +\mu_0 N+\frac{1}{2}M v_{\s}^2 - P_c v_{\s},\nonumber\\
    F^\mathrm{SF} &= F - P_c v_{\s}.
\end{align}
Free energy considered above is free energy of the whole system, with or without the soliton ($F_\s$ or $F_0$). We are interested in the free energy of the soliton itself:
\begin{align}
    \Delta F =F_\s-F_0.
\end{align}
Knowing that $P_0=0$ and $P_\s=M_\mathrm{d} v_\mathrm{B}$ where $M_\mathrm{d}$ is depletion mass ($M_\mathrm{d}=m(N_\s-N_0)$), we get:
\begin{align}\label{eq:Ftransform}
    \Delta F =\Delta F_{v_\s}+ M_\mathrm{d} v_\s^2.
\end{align}

\bibliography{bibl}

\begin{thebibliography}{48}%
\makeatletter
\providecommand \@ifxundefined [1]{%
 \@ifx{#1\undefined}
}%
\providecommand \@ifnum [1]{%
 \ifnum #1\expandafter \@firstoftwo
 \else \expandafter \@secondoftwo
 \fi
}%
\providecommand \@ifx [1]{%
 \ifx #1\expandafter \@firstoftwo
 \else \expandafter \@secondoftwo
 \fi
}%
\providecommand \natexlab [1]{#1}%
\providecommand \enquote  [1]{``#1''}%
\providecommand \bibnamefont  [1]{#1}%
\providecommand \bibfnamefont [1]{#1}%
\providecommand \citenamefont [1]{#1}%
\providecommand \href@noop [0]{\@secondoftwo}%
\providecommand \href [0]{\begingroup \@sanitize@url \@href}%
\providecommand \@href[1]{\@@startlink{#1}\@@href}%
\providecommand \@@href[1]{\endgroup#1\@@endlink}%
\providecommand \@sanitize@url [0]{\catcode `\\12\catcode `\$12\catcode `\&12\catcode `\#12\catcode `\^12\catcode `\_12\catcode `\%12\relax}%
\providecommand \@@startlink[1]{}%
\providecommand \@@endlink[0]{}%
\providecommand \url  [0]{\begingroup\@sanitize@url \@url }%
\providecommand \@url [1]{\endgroup\@href {#1}{\urlprefix }}%
\providecommand \urlprefix  [0]{URL }%
\providecommand \Eprint [0]{\href }%
\providecommand \doibase [0]{https://doi.org/}%
\providecommand \selectlanguage [0]{\@gobble}%
\providecommand \bibinfo  [0]{\@secondoftwo}%
\providecommand \bibfield  [0]{\@secondoftwo}%
\providecommand \translation [1]{[#1]}%
\providecommand \BibitemOpen [0]{}%
\providecommand \bibitemStop [0]{}%
\providecommand \bibitemNoStop [0]{.\EOS\space}%
\providecommand \EOS [0]{\spacefactor3000\relax}%
\providecommand \BibitemShut  [1]{\csname bibitem#1\endcsname}%
\let\auto@bib@innerbib\@empty
\bibitem [{\citenamefont {Onnes}(1911)}]{Onnes1911}%
  \BibitemOpen
  \bibfield  {author} {\bibinfo {author} {\bibfnamefont {H.~K.}\ \bibnamefont {Onnes}},\ }\bibfield  {title} {\bibinfo {title} {The resistance of pure mercury at helium temperatures},\ }\href@noop {} {\bibfield  {journal} {\bibinfo  {journal} {Commun. Phys. Lab. Univ. Leiden. Suppl.}\ }\textbf {\bibinfo {volume} {29}} (\bibinfo {year} {1911})}\BibitemShut {NoStop}%
\bibitem [{\citenamefont {Kapitza}(1938)}]{Kapitza1938}%
  \BibitemOpen
  \bibfield  {author} {\bibinfo {author} {\bibfnamefont {P.}~\bibnamefont {Kapitza}},\ }\bibfield  {title} {\bibinfo {title} {Viscosity of {{Liquid Helium}} below the {$\lambda$}-{{Point}}},\ }\href {https://doi.org/10.1038/141074a0} {\bibfield  {journal} {\bibinfo  {journal} {Nature}\ }\textbf {\bibinfo {volume} {141}},\ \bibinfo {pages} {74} (\bibinfo {year} {1938})}\BibitemShut {NoStop}%
\bibitem [{\citenamefont {Allen}\ and\ \citenamefont {Misener}(1938)}]{Allen1938}%
  \BibitemOpen
  \bibfield  {author} {\bibinfo {author} {\bibfnamefont {J.~F.}\ \bibnamefont {Allen}}\ and\ \bibinfo {author} {\bibfnamefont {A.~D.}\ \bibnamefont {Misener}},\ }\bibfield  {title} {\bibinfo {title} {Flow {{Phenomena}} in {{Liquid Helium II}}},\ }\href {https://doi.org/10.1038/142643a0} {\bibfield  {journal} {\bibinfo  {journal} {Nature}\ }\textbf {\bibinfo {volume} {142}},\ \bibinfo {pages} {643} (\bibinfo {year} {1938})}\BibitemShut {NoStop}%
\bibitem [{\citenamefont {Korteweg}\ and\ \citenamefont {de~Vries}(1895)}]{Korteweg1895}%
  \BibitemOpen
  \bibfield  {author} {\bibinfo {author} {\bibfnamefont {D.~D.~J.}\ \bibnamefont {Korteweg}}\ and\ \bibinfo {author} {\bibfnamefont {D.~G.}\ \bibnamefont {de~Vries}},\ }\bibfield  {title} {\bibinfo {title} {Xli. on the change of form of long waves advancing in a rectangular canal, and on a new type of long stationary waves},\ }\href {https://doi.org/10.1080/14786449508620739} {\bibfield  {journal} {\bibinfo  {journal} {The London, Edinburgh, and Dublin Philosophical Magazine and Journal of Science}\ }\textbf {\bibinfo {volume} {39}},\ \bibinfo {pages} {422} (\bibinfo {year} {1895})},\ \Eprint {https://arxiv.org/abs/https://doi.org/10.1080/14786449508620739} {https://doi.org/10.1080/14786449508620739} \BibitemShut {NoStop}%
\bibitem [{\citenamefont {Muto}\ \emph {et~al.}(1988)\citenamefont {Muto}, \citenamefont {Halding}, \citenamefont {Christiansen},\ and\ \citenamefont {Scott}}]{Muto1988}%
  \BibitemOpen
  \bibfield  {author} {\bibinfo {author} {\bibfnamefont {V.}~\bibnamefont {Muto}}, \bibinfo {author} {\bibfnamefont {J.}~\bibnamefont {Halding}}, \bibinfo {author} {\bibfnamefont {P.}~\bibnamefont {Christiansen}},\ and\ \bibinfo {author} {\bibfnamefont {A.}~\bibnamefont {Scott}},\ }\bibfield  {title} {\bibinfo {title} {Solitons in {DNA}},\ }\href {https://doi.org/10.1080/07391102.1988.10506432} {\bibfield  {journal} {\bibinfo  {journal} {J Biomol Struct Dyn.}\ }\textbf {\bibinfo {volume} {5}},\ \bibinfo {pages} {873} (\bibinfo {year} {1988})}\BibitemShut {NoStop}%
\bibitem [{\citenamefont {Zurek}(1996)}]{ZUREK1996}%
  \BibitemOpen
  \bibfield  {author} {\bibinfo {author} {\bibfnamefont {W.}~\bibnamefont {Zurek}},\ }\bibfield  {title} {\bibinfo {title} {Cosmological experiments in condensed matter systems},\ }\href {https://doi.org/https://doi.org/10.1016/S0370-1573(96)00009-9} {\bibfield  {journal} {\bibinfo  {journal} {Physics Reports}\ }\textbf {\bibinfo {volume} {276}},\ \bibinfo {pages} {177} (\bibinfo {year} {1996})}\BibitemShut {NoStop}%
\bibitem [{\citenamefont {Tsuzuki}(1971)}]{Tsuzuki1971}%
  \BibitemOpen
  \bibfield  {author} {\bibinfo {author} {\bibfnamefont {T.}~\bibnamefont {Tsuzuki}},\ }\bibfield  {title} {\bibinfo {title} {Nonlinear waves in the {{Pitaevskii-Gross}} equation},\ }\href {https://doi.org/10.1007/BF00628744} {\bibfield  {journal} {\bibinfo  {journal} {J. Low Temp. Phys.}\ }\textbf {\bibinfo {volume} {4}},\ \bibinfo {pages} {441} (\bibinfo {year} {1971})}\BibitemShut {NoStop}%
\bibitem [{\citenamefont {Burger}\ \emph {et~al.}(1999)\citenamefont {Burger}, \citenamefont {Bongs}, \citenamefont {Dettmer}, \citenamefont {Ertmer}, \citenamefont {Sengstock}, \citenamefont {Sanpera}, \citenamefont {Shlyapnikov},\ and\ \citenamefont {Lewenstein}}]{Burger1999}%
  \BibitemOpen
  \bibfield  {author} {\bibinfo {author} {\bibfnamefont {S.}~\bibnamefont {Burger}}, \bibinfo {author} {\bibfnamefont {K.}~\bibnamefont {Bongs}}, \bibinfo {author} {\bibfnamefont {S.}~\bibnamefont {Dettmer}}, \bibinfo {author} {\bibfnamefont {W.}~\bibnamefont {Ertmer}}, \bibinfo {author} {\bibfnamefont {K.}~\bibnamefont {Sengstock}}, \bibinfo {author} {\bibfnamefont {A.}~\bibnamefont {Sanpera}}, \bibinfo {author} {\bibfnamefont {G.~V.}\ \bibnamefont {Shlyapnikov}},\ and\ \bibinfo {author} {\bibfnamefont {M.}~\bibnamefont {Lewenstein}},\ }\bibfield  {title} {\bibinfo {title} {Dark solitons in {B}ose-{E}instein condensates},\ }\href {https://doi.org/10.1103/PhysRevLett.83.5198} {\bibfield  {journal} {\bibinfo  {journal} {Phys. Rev. Lett.}\ }\textbf {\bibinfo {volume} {83}},\ \bibinfo {pages} {5198} (\bibinfo {year} {1999})}\BibitemShut {NoStop}%
\bibitem [{\citenamefont {Denschlag}\ \emph {et~al.}(2000)\citenamefont {Denschlag}, \citenamefont {Simsarian}, \citenamefont {Feder}, \citenamefont {Clark}, \citenamefont {Collins}, \citenamefont {Cubizolles}, \citenamefont {Deng}, \citenamefont {Hagley}, \citenamefont {Helmerson}, \citenamefont {Reinhardt}, \citenamefont {Rolston}, \citenamefont {Schneider},\ and\ \citenamefont {Phillips}}]{Denschlag2000}%
  \BibitemOpen
  \bibfield  {author} {\bibinfo {author} {\bibfnamefont {J.}~\bibnamefont {Denschlag}}, \bibinfo {author} {\bibfnamefont {J.~E.}\ \bibnamefont {Simsarian}}, \bibinfo {author} {\bibfnamefont {D.~L.}\ \bibnamefont {Feder}}, \bibinfo {author} {\bibfnamefont {C.~W.}\ \bibnamefont {Clark}}, \bibinfo {author} {\bibfnamefont {L.~A.}\ \bibnamefont {Collins}}, \bibinfo {author} {\bibfnamefont {J.}~\bibnamefont {Cubizolles}}, \bibinfo {author} {\bibfnamefont {L.}~\bibnamefont {Deng}}, \bibinfo {author} {\bibfnamefont {E.~W.}\ \bibnamefont {Hagley}}, \bibinfo {author} {\bibfnamefont {K.}~\bibnamefont {Helmerson}}, \bibinfo {author} {\bibfnamefont {W.~P.}\ \bibnamefont {Reinhardt}}, \bibinfo {author} {\bibfnamefont {S.~L.}\ \bibnamefont {Rolston}}, \bibinfo {author} {\bibfnamefont {B.~I.}\ \bibnamefont {Schneider}},\ and\ \bibinfo {author} {\bibfnamefont {W.~D.}\ \bibnamefont {Phillips}},\ }\bibfield  {title} {\bibinfo {title} {Generating {{Solitons}} by {{Phase Engineering}} of a {{Bose-Einstein Condensate}}},\ }\href
  {https://doi.org/10.1126/science.287.5450.97} {\bibfield  {journal} {\bibinfo  {journal} {Science}\ }\textbf {\bibinfo {volume} {287}},\ \bibinfo {pages} {97} (\bibinfo {year} {2000})}\BibitemShut {NoStop}%
\bibitem [{\citenamefont {Becker}\ \emph {et~al.}(2008)\citenamefont {Becker}, \citenamefont {Stellmer}, \citenamefont {{Soltan-Panahi}}, \citenamefont {D{\"o}rscher}, \citenamefont {Baumert}, \citenamefont {Richter}, \citenamefont {Kronj{\"a}ger}, \citenamefont {Bongs},\ and\ \citenamefont {Sengstock}}]{Becker2008}%
  \BibitemOpen
  \bibfield  {author} {\bibinfo {author} {\bibfnamefont {C.}~\bibnamefont {Becker}}, \bibinfo {author} {\bibfnamefont {S.}~\bibnamefont {Stellmer}}, \bibinfo {author} {\bibfnamefont {P.}~\bibnamefont {{Soltan-Panahi}}}, \bibinfo {author} {\bibfnamefont {S.}~\bibnamefont {D{\"o}rscher}}, \bibinfo {author} {\bibfnamefont {M.}~\bibnamefont {Baumert}}, \bibinfo {author} {\bibfnamefont {E.-M.}\ \bibnamefont {Richter}}, \bibinfo {author} {\bibfnamefont {J.}~\bibnamefont {Kronj{\"a}ger}}, \bibinfo {author} {\bibfnamefont {K.}~\bibnamefont {Bongs}},\ and\ \bibinfo {author} {\bibfnamefont {K.}~\bibnamefont {Sengstock}},\ }\bibfield  {title} {\bibinfo {title} {Oscillations and interactions of dark and dark--bright solitons in {{Bose}}--{{Einstein}} condensates},\ }\href {https://doi.org/10.1038/nphys962} {\bibfield  {journal} {\bibinfo  {journal} {Nat. Phys.}\ }\textbf {\bibinfo {volume} {4}},\ \bibinfo {pages} {496} (\bibinfo {year} {2008})}\BibitemShut {NoStop}%
\bibitem [{\citenamefont {Ku}\ \emph {et~al.}(2016)\citenamefont {Ku}, \citenamefont {Mukherjee}, \citenamefont {Yefsah},\ and\ \citenamefont {Zwierlein}}]{Ku2016}%
  \BibitemOpen
  \bibfield  {author} {\bibinfo {author} {\bibfnamefont {M.~J.~H.}\ \bibnamefont {Ku}}, \bibinfo {author} {\bibfnamefont {B.}~\bibnamefont {Mukherjee}}, \bibinfo {author} {\bibfnamefont {T.}~\bibnamefont {Yefsah}},\ and\ \bibinfo {author} {\bibfnamefont {M.~W.}\ \bibnamefont {Zwierlein}},\ }\bibfield  {title} {\bibinfo {title} {Cascade of {{Solitonic Excitations}} in a {{Superfluid Fermi}} gas: {{From Planar Solitons}} to {{Vortex Rings}} and {{Lines}}},\ }\href {https://doi.org/10.1103/PhysRevLett.116.045304} {\bibfield  {journal} {\bibinfo  {journal} {Phys. Rev. Lett.}\ }\textbf {\bibinfo {volume} {116}},\ \bibinfo {pages} {045304} (\bibinfo {year} {2016})},\ \Eprint {https://arxiv.org/abs/1507.01047} {arxiv:1507.01047} \BibitemShut {NoStop}%
\bibitem [{\citenamefont {Shamailov}\ and\ \citenamefont {Brand}(2016)}]{Shamailov2016}%
  \BibitemOpen
  \bibfield  {author} {\bibinfo {author} {\bibfnamefont {S.~S.}\ \bibnamefont {Shamailov}}\ and\ \bibinfo {author} {\bibfnamefont {J.}~\bibnamefont {Brand}},\ }\bibfield  {title} {\bibinfo {title} {Dark-soliton-like excitations in the {Y}ang{\textendash}{G}audin gas of attractively interacting fermions},\ }\href {https://doi.org/10.1088/1367-2630/18/7/075004} {\bibfield  {journal} {\bibinfo  {journal} {New Journal of Physics}\ }\textbf {\bibinfo {volume} {18}},\ \bibinfo {pages} {075004} (\bibinfo {year} {2016})}\BibitemShut {NoStop}%
\bibitem [{\citenamefont {Efimkin}\ and\ \citenamefont {Galitski}(2015)}]{Efimkin2015}%
  \BibitemOpen
  \bibfield  {author} {\bibinfo {author} {\bibfnamefont {D.~K.}\ \bibnamefont {Efimkin}}\ and\ \bibinfo {author} {\bibfnamefont {V.}~\bibnamefont {Galitski}},\ }\bibfield  {title} {\bibinfo {title} {Moving solitons in a one-dimensional fermionic superfluid},\ }\href {https://doi.org/10.1103/PhysRevA.91.023616} {\bibfield  {journal} {\bibinfo  {journal} {Phys. Rev. A}\ }\textbf {\bibinfo {volume} {91}},\ \bibinfo {pages} {023616} (\bibinfo {year} {2015})}\BibitemShut {NoStop}%
\bibitem [{\citenamefont {Zou}\ \emph {et~al.}(2016)\citenamefont {Zou}, \citenamefont {Brand}, \citenamefont {Liu},\ and\ \citenamefont {Hu}}]{Zou2016}%
  \BibitemOpen
  \bibfield  {author} {\bibinfo {author} {\bibfnamefont {P.}~\bibnamefont {Zou}}, \bibinfo {author} {\bibfnamefont {J.}~\bibnamefont {Brand}}, \bibinfo {author} {\bibfnamefont {X.-J.}\ \bibnamefont {Liu}},\ and\ \bibinfo {author} {\bibfnamefont {H.}~\bibnamefont {Hu}},\ }\bibfield  {title} {\bibinfo {title} {Traveling {{Majorana Solitons}} in a {{Low-Dimensional Spin-Orbit-Coupled Fermi Superfluid}}},\ }\href {https://doi.org/10.1103/PhysRevLett.117.225302} {\bibfield  {journal} {\bibinfo  {journal} {Phys. Rev. Lett.}\ }\textbf {\bibinfo {volume} {117}},\ \bibinfo {pages} {225302} (\bibinfo {year} {2016})},\ \Eprint {https://arxiv.org/abs/1509.01803} {arxiv:1509.01803} \BibitemShut {NoStop}%
\bibitem [{\citenamefont {Dziarmaga}\ and\ \citenamefont {Sacha}(2004)}]{Dziarmaga2004c}%
  \BibitemOpen
  \bibfield  {author} {\bibinfo {author} {\bibfnamefont {J.}~\bibnamefont {Dziarmaga}}\ and\ \bibinfo {author} {\bibfnamefont {K.}~\bibnamefont {Sacha}},\ }\href {https://doi.org/10.48550/arXiv.cond-mat/0407585} {\bibinfo {title} {Soliton in {{BCS}} superfluid {{Fermi}} gas}} (\bibinfo {year} {2004}),\ \Eprint {https://arxiv.org/abs/cond-mat/0407585} {arxiv:cond-mat/0407585} \BibitemShut {NoStop}%
\bibitem [{\citenamefont {Antezza}\ \emph {et~al.}(2007)\citenamefont {Antezza}, \citenamefont {Dalfovo}, \citenamefont {Pitaevskii},\ and\ \citenamefont {Stringari}}]{Antezza2007}%
  \BibitemOpen
  \bibfield  {author} {\bibinfo {author} {\bibfnamefont {M.}~\bibnamefont {Antezza}}, \bibinfo {author} {\bibfnamefont {F.}~\bibnamefont {Dalfovo}}, \bibinfo {author} {\bibfnamefont {L.~P.}\ \bibnamefont {Pitaevskii}},\ and\ \bibinfo {author} {\bibfnamefont {S.}~\bibnamefont {Stringari}},\ }\bibfield  {title} {\bibinfo {title} {Dark solitons in a superfluid {F}ermi gas},\ }\href {https://doi.org/10.1103/PhysRevA.76.043610} {\bibfield  {journal} {\bibinfo  {journal} {Phys. Rev. A}\ }\textbf {\bibinfo {volume} {76}},\ \bibinfo {pages} {043610} (\bibinfo {year} {2007})}\BibitemShut {NoStop}%
\bibitem [{\citenamefont {Liao}\ and\ \citenamefont {Brand}(2011)}]{Liao11pr:FermiSolitons}%
  \BibitemOpen
  \bibfield  {author} {\bibinfo {author} {\bibfnamefont {R.}~\bibnamefont {Liao}}\ and\ \bibinfo {author} {\bibfnamefont {J.}~\bibnamefont {Brand}},\ }\bibfield  {title} {\bibinfo {title} {Traveling dark solitons in superfluid {{Fermi}} gases},\ }\href {https://doi.org/10.1103/PhysRevA.83.041604} {\bibfield  {journal} {\bibinfo  {journal} {Phys. Rev. A}\ }\textbf {\bibinfo {volume} {83}},\ \bibinfo {pages} {041604(R)} (\bibinfo {year} {2011})}\BibitemShut {NoStop}%
\bibitem [{\citenamefont {Scott}\ \emph {et~al.}(2011)\citenamefont {Scott}, \citenamefont {Dalfovo}, \citenamefont {Pitaevskii},\ and\ \citenamefont {Stringari}}]{Scott2011}%
  \BibitemOpen
  \bibfield  {author} {\bibinfo {author} {\bibfnamefont {R.}~\bibnamefont {Scott}}, \bibinfo {author} {\bibfnamefont {F.}~\bibnamefont {Dalfovo}}, \bibinfo {author} {\bibfnamefont {L.}~\bibnamefont {Pitaevskii}},\ and\ \bibinfo {author} {\bibfnamefont {S.}~\bibnamefont {Stringari}},\ }\bibfield  {title} {\bibinfo {title} {Dynamics of {{Dark Solitons}} in a {{Trapped Superfluid Fermi Gas}}},\ }\href {https://doi.org/10.1103/PhysRevLett.106.185301} {\bibfield  {journal} {\bibinfo  {journal} {Phys. Rev. Lett.}\ }\textbf {\bibinfo {volume} {106}},\ \bibinfo {pages} {185301} (\bibinfo {year} {2011})}\BibitemShut {NoStop}%
\bibitem [{\citenamefont {Spuntarelli}\ \emph {et~al.}(2011)\citenamefont {Spuntarelli}, \citenamefont {Carr}, \citenamefont {Pieri},\ and\ \citenamefont {Strinati}}]{Spuntarelli2011}%
  \BibitemOpen
  \bibfield  {author} {\bibinfo {author} {\bibfnamefont {A.}~\bibnamefont {Spuntarelli}}, \bibinfo {author} {\bibfnamefont {L.~D.}\ \bibnamefont {Carr}}, \bibinfo {author} {\bibfnamefont {P.}~\bibnamefont {Pieri}},\ and\ \bibinfo {author} {\bibfnamefont {G.~C.}\ \bibnamefont {Strinati}},\ }\bibfield  {title} {\bibinfo {title} {Gray solitons in a strongly interacting superfluid {{Fermi}} gas},\ }\href {https://doi.org/10.1088/1367-2630/13/3/035010} {\bibfield  {journal} {\bibinfo  {journal} {New J. Phys.}\ }\textbf {\bibinfo {volume} {13}},\ \bibinfo {pages} {035010} (\bibinfo {year} {2011})}\BibitemShut {NoStop}%
\bibitem [{\citenamefont {Scott}\ \emph {et~al.}(2012)\citenamefont {Scott}, \citenamefont {Dalfovo}, \citenamefont {Pitaevskii}, \citenamefont {Stringari}, \citenamefont {Fialko}, \citenamefont {Liao},\ and\ \citenamefont {Brand}}]{Scott2012}%
  \BibitemOpen
  \bibfield  {author} {\bibinfo {author} {\bibfnamefont {R.~G.}\ \bibnamefont {Scott}}, \bibinfo {author} {\bibfnamefont {F.}~\bibnamefont {Dalfovo}}, \bibinfo {author} {\bibfnamefont {L.~P.}\ \bibnamefont {Pitaevskii}}, \bibinfo {author} {\bibfnamefont {S.}~\bibnamefont {Stringari}}, \bibinfo {author} {\bibfnamefont {O.}~\bibnamefont {Fialko}}, \bibinfo {author} {\bibfnamefont {R.}~\bibnamefont {Liao}},\ and\ \bibinfo {author} {\bibfnamefont {J.}~\bibnamefont {Brand}},\ }\bibfield  {title} {\bibinfo {title} {The decay and collisions of dark solitons in superfluid {{Fermi}} gases},\ }\href {https://doi.org/10.1088/1367-2630/14/2/023044} {\bibfield  {journal} {\bibinfo  {journal} {New J. Phys.}\ }\textbf {\bibinfo {volume} {14}},\ \bibinfo {pages} {023044} (\bibinfo {year} {2012})},\ \Eprint {https://arxiv.org/abs/1109.6444} {arxiv:1109.6444} \BibitemShut {NoStop}%
\bibitem [{\citenamefont {Klimin}\ \emph {et~al.}(2014)\citenamefont {Klimin}, \citenamefont {Tempere},\ and\ \citenamefont {Devreese}}]{Klimin2014}%
  \BibitemOpen
  \bibfield  {author} {\bibinfo {author} {\bibfnamefont {S.~N.}\ \bibnamefont {Klimin}}, \bibinfo {author} {\bibfnamefont {J.}~\bibnamefont {Tempere}},\ and\ \bibinfo {author} {\bibfnamefont {J.~T.}\ \bibnamefont {Devreese}},\ }\bibfield  {title} {\bibinfo {title} {Finite-temperature effective field theory for dark solitons in superfluid {{Fermi}} gases},\ }\href {https://doi.org/10.1103/PhysRevA.90.053613} {\bibfield  {journal} {\bibinfo  {journal} {Phys. Rev. A}\ }\textbf {\bibinfo {volume} {90}},\ \bibinfo {pages} {053613} (\bibinfo {year} {2014})}\BibitemShut {NoStop}%
\bibitem [{\citenamefont {Van~Alphen}\ \emph {et~al.}(2019)\citenamefont {Van~Alphen}, \citenamefont {Takeuchi},\ and\ \citenamefont {Tempere}}]{VanAlphen2019a}%
  \BibitemOpen
  \bibfield  {author} {\bibinfo {author} {\bibfnamefont {W.}~\bibnamefont {Van~Alphen}}, \bibinfo {author} {\bibfnamefont {H.}~\bibnamefont {Takeuchi}},\ and\ \bibinfo {author} {\bibfnamefont {J.}~\bibnamefont {Tempere}},\ }\bibfield  {title} {\bibinfo {title} {Crossover between snake instability and {{Josephson}} instability of dark solitons in superfluid {{Fermi}} gases},\ }\href {https://doi.org/10.1103/PhysRevA.100.023628} {\bibfield  {journal} {\bibinfo  {journal} {Phys. Rev. A}\ }\textbf {\bibinfo {volume} {100}},\ \bibinfo {pages} {023628} (\bibinfo {year} {2019})}\BibitemShut {NoStop}%
\bibitem [{\citenamefont {Keimer}\ \emph {et~al.}(2015)\citenamefont {Keimer}, \citenamefont {Kivelson}, \citenamefont {Norman}, \citenamefont {Uchida},\ and\ \citenamefont {Zaanen}}]{Keimer2015}%
  \BibitemOpen
  \bibfield  {author} {\bibinfo {author} {\bibfnamefont {B.}~\bibnamefont {Keimer}}, \bibinfo {author} {\bibfnamefont {S.~A.}\ \bibnamefont {Kivelson}}, \bibinfo {author} {\bibfnamefont {M.~R.}\ \bibnamefont {Norman}}, \bibinfo {author} {\bibfnamefont {S.}~\bibnamefont {Uchida}},\ and\ \bibinfo {author} {\bibfnamefont {J.}~\bibnamefont {Zaanen}},\ }\bibfield  {title} {\bibinfo {title} {From quantum matter to high-temperature superconductivity in copper oxides},\ }\href {https://doi.org/10.1038/nature14165} {\bibfield  {journal} {\bibinfo  {journal} {Nature}\ }\textbf {\bibinfo {volume} {518}},\ \bibinfo {pages} {179} (\bibinfo {year} {2015})}\BibitemShut {NoStop}%
\bibitem [{\citenamefont {Sato}\ and\ \citenamefont {Ando}(2017)}]{Sato2017}%
  \BibitemOpen
  \bibfield  {author} {\bibinfo {author} {\bibfnamefont {M.}~\bibnamefont {Sato}}\ and\ \bibinfo {author} {\bibfnamefont {Y.}~\bibnamefont {Ando}},\ }\bibfield  {title} {\bibinfo {title} {Topological superconductors: a review},\ }\href {https://doi.org/10.1088/1361-6633/aa6ac7} {\bibfield  {journal} {\bibinfo  {journal} {Reports on Progress in Physics}\ }\textbf {\bibinfo {volume} {80}},\ \bibinfo {pages} {076501} (\bibinfo {year} {2017})}\BibitemShut {NoStop}%
\bibitem [{\citenamefont {Challis}\ \emph {et~al.}(2007)\citenamefont {Challis}, \citenamefont {Ballagh},\ and\ \citenamefont {Gardiner}}]{Challis2007}%
  \BibitemOpen
  \bibfield  {author} {\bibinfo {author} {\bibfnamefont {K.~J.}\ \bibnamefont {Challis}}, \bibinfo {author} {\bibfnamefont {R.~J.}\ \bibnamefont {Ballagh}},\ and\ \bibinfo {author} {\bibfnamefont {C.~W.}\ \bibnamefont {Gardiner}},\ }\bibfield  {title} {\bibinfo {title} {Bragg scattering of cooper pairs in an ultracold fermi gas},\ }\href {https://doi.org/10.1103/PhysRevLett.98.093002} {\bibfield  {journal} {\bibinfo  {journal} {Phys. Rev. Lett.}\ }\textbf {\bibinfo {volume} {98}},\ \bibinfo {pages} {093002} (\bibinfo {year} {2007})}\BibitemShut {NoStop}%
\bibitem [{\citenamefont {Bloch}\ \emph {et~al.}(2012)\citenamefont {Bloch}, \citenamefont {Dalibard},\ and\ \citenamefont {Nascimb{\`e}ne}}]{Bloch2012}%
  \BibitemOpen
  \bibfield  {author} {\bibinfo {author} {\bibfnamefont {I.}~\bibnamefont {Bloch}}, \bibinfo {author} {\bibfnamefont {J.}~\bibnamefont {Dalibard}},\ and\ \bibinfo {author} {\bibfnamefont {S.}~\bibnamefont {Nascimb{\`e}ne}},\ }\bibfield  {title} {\bibinfo {title} {Quantum simulations with ultracold quantum gases},\ }\href {https://doi.org/10.1038/nphys2259} {\bibfield  {journal} {\bibinfo  {journal} {Nature Physics}\ }\textbf {\bibinfo {volume} {8}},\ \bibinfo {pages} {267} (\bibinfo {year} {2012})}\BibitemShut {NoStop}%
\bibitem [{\citenamefont {Chin}\ \emph {et~al.}(2010)\citenamefont {Chin}, \citenamefont {Grimm}, \citenamefont {Julienne},\ and\ \citenamefont {Tiesinga}}]{Chin2010}%
  \BibitemOpen
  \bibfield  {author} {\bibinfo {author} {\bibfnamefont {C.}~\bibnamefont {Chin}}, \bibinfo {author} {\bibfnamefont {R.}~\bibnamefont {Grimm}}, \bibinfo {author} {\bibfnamefont {P.}~\bibnamefont {Julienne}},\ and\ \bibinfo {author} {\bibfnamefont {E.}~\bibnamefont {Tiesinga}},\ }\bibfield  {title} {\bibinfo {title} {Feshbach resonances in ultracold gases},\ }\href {https://doi.org/10.1103/RevModPhys.82.1225} {\bibfield  {journal} {\bibinfo  {journal} {Rev. Mod. Phys.}\ }\textbf {\bibinfo {volume} {82}},\ \bibinfo {pages} {1225} (\bibinfo {year} {2010})}\BibitemShut {NoStop}%
\bibitem [{\citenamefont {Bulgac}\ \emph {et~al.}(2011)\citenamefont {Bulgac}, \citenamefont {Luo}, \citenamefont {Magierski}, \citenamefont {Roche},\ and\ \citenamefont {Yu}}]{Bulgac2011}%
  \BibitemOpen
  \bibfield  {author} {\bibinfo {author} {\bibfnamefont {A.}~\bibnamefont {Bulgac}}, \bibinfo {author} {\bibfnamefont {Y.-L.}\ \bibnamefont {Luo}}, \bibinfo {author} {\bibfnamefont {P.}~\bibnamefont {Magierski}}, \bibinfo {author} {\bibfnamefont {K.~J.}\ \bibnamefont {Roche}},\ and\ \bibinfo {author} {\bibfnamefont {Y.}~\bibnamefont {Yu}},\ }\bibfield  {title} {\bibinfo {title} {Real-{{Time Dynamics}} of {{Quantized Vortices}} in a {{Unitary Fermi Superfluid}}},\ }\href {https://doi.org/10.1126/science.1201968} {\bibfield  {journal} {\bibinfo  {journal} {Science}\ }\textbf {\bibinfo {volume} {332}},\ \bibinfo {pages} {1288} (\bibinfo {year} {2011})}\BibitemShut {NoStop}%
\bibitem [{\citenamefont {Bertaina}\ and\ \citenamefont {Giorgini}(2011)}]{Bertaina2011}%
  \BibitemOpen
  \bibfield  {author} {\bibinfo {author} {\bibfnamefont {G.}~\bibnamefont {Bertaina}}\ and\ \bibinfo {author} {\bibfnamefont {S.}~\bibnamefont {Giorgini}},\ }\bibfield  {title} {\bibinfo {title} {Bcs-bec crossover in a two-dimensional fermi gas},\ }\href {https://doi.org/10.1103/PhysRevLett.106.110403} {\bibfield  {journal} {\bibinfo  {journal} {Phys. Rev. Lett.}\ }\textbf {\bibinfo {volume} {106}},\ \bibinfo {pages} {110403} (\bibinfo {year} {2011})}\BibitemShut {NoStop}%
\bibitem [{\citenamefont {De~Gennes}(1999)}]{DeGennes1999}%
  \BibitemOpen
  \bibfield  {author} {\bibinfo {author} {\bibfnamefont {P.~G.}\ \bibnamefont {De~Gennes}},\ }\href {https://doi.org/https://doi.org/10.1201/9780429497032} {\emph {\bibinfo {title} {Superconductivity of Metals and Alloys}}}\ (\bibinfo  {publisher} {CRC Press},\ \bibinfo {year} {1999})\BibitemShut {NoStop}%
\bibitem [{Note1()}]{Note1}%
  \BibitemOpen
  \bibinfo {note} {Due to $\Delta (x)$ being a sum of products of the eigenstates of the BdG matrix, phase twists applied to either diagonal sector of BdG matrix ($\phi _u$ and $\phi _v$, respectively) add in contributing to $\Delta (x)$, while the individual phases do not have physical significance. In our calculations we usually set $\phi _u$ to $0$ and $\phi _v$ to $\pi $.}\BibitemShut {Stop}%
\bibitem [{\citenamefont {Shamailov}\ and\ \citenamefont {Brand}(2019)}]{Shamailov2019a}%
  \BibitemOpen
  \bibfield  {author} {\bibinfo {author} {\bibfnamefont {S.~S.}\ \bibnamefont {Shamailov}}\ and\ \bibinfo {author} {\bibfnamefont {J.}~\bibnamefont {Brand}},\ }\bibfield  {title} {\bibinfo {title} {Quantum dark solitons in the one-dimensional {{Bose}} gas},\ }\href {https://doi.org/10.1103/PhysRevA.99.043632} {\bibfield  {journal} {\bibinfo  {journal} {Phys. Rev. A}\ }\textbf {\bibinfo {volume} {99}},\ \bibinfo {pages} {043632} (\bibinfo {year} {2019})}\BibitemShut {NoStop}%
\bibitem [{\citenamefont {Konotop}\ and\ \citenamefont {Pitaevskii}(2004)}]{Konotop2004}%
  \BibitemOpen
  \bibfield  {author} {\bibinfo {author} {\bibfnamefont {V.~V.}\ \bibnamefont {Konotop}}\ and\ \bibinfo {author} {\bibfnamefont {L.}~\bibnamefont {Pitaevskii}},\ }\bibfield  {title} {\bibinfo {title} {Landau {{Dynamics}} of a {{Grey Soliton}} in a {{Trapped Condensate}}},\ }\href {https://doi.org/10.1103/PhysRevLett.93.240403} {\bibfield  {journal} {\bibinfo  {journal} {Phys. Rev. Lett.}\ }\textbf {\bibinfo {volume} {93}},\ \bibinfo {pages} {240403} (\bibinfo {year} {2004})}\BibitemShut {NoStop}%
\bibitem [{\citenamefont {Mateo}\ and\ \citenamefont {Brand}(2015)}]{Mateo2015}%
  \BibitemOpen
  \bibfield  {author} {\bibinfo {author} {\bibfnamefont {A.~M.}\ \bibnamefont {Mateo}}\ and\ \bibinfo {author} {\bibfnamefont {J.}~\bibnamefont {Brand}},\ }\bibfield  {title} {\bibinfo {title} {Stability and dispersion relations of three-dimensional solitary waves in trapped {B}ose{\textendash}{E}instein condensates},\ }\href {https://doi.org/10.1088/1367-2630/17/12/125013} {\bibfield  {journal} {\bibinfo  {journal} {New Journal of Physics}\ }\textbf {\bibinfo {volume} {17}},\ \bibinfo {pages} {125013} (\bibinfo {year} {2015})}\BibitemShut {NoStop}%
\bibitem [{\citenamefont {Randeria}\ \emph {et~al.}(1989)\citenamefont {Randeria}, \citenamefont {Duan},\ and\ \citenamefont {Shieh}}]{Randeria1989}%
  \BibitemOpen
  \bibfield  {author} {\bibinfo {author} {\bibfnamefont {M.}~\bibnamefont {Randeria}}, \bibinfo {author} {\bibfnamefont {J.-M.}\ \bibnamefont {Duan}},\ and\ \bibinfo {author} {\bibfnamefont {L.-Y.}\ \bibnamefont {Shieh}},\ }\bibfield  {title} {\bibinfo {title} {Bound states, cooper pairing, and bose condensation in two dimensions},\ }\href {https://doi.org/10.1103/PhysRevLett.62.981} {\bibfield  {journal} {\bibinfo  {journal} {Phys. Rev. Lett.}\ }\textbf {\bibinfo {volume} {62}},\ \bibinfo {pages} {981} (\bibinfo {year} {1989})}\BibitemShut {NoStop}%
\bibitem [{\citenamefont {Broyden}(1965)}]{Broyden1965}%
  \BibitemOpen
  \bibfield  {author} {\bibinfo {author} {\bibfnamefont {C.~G.}\ \bibnamefont {Broyden}},\ }\bibfield  {title} {\bibinfo {title} {A class of methods for solving nonlinear simultaneous equations},\ }\href {https://doi.org/10.1090/S0025-5718-1965-0198670-6} {\bibfield  {journal} {\bibinfo  {journal} {Math. Comp.}\ }\textbf {\bibinfo {volume} {19}},\ \bibinfo {pages} {577} (\bibinfo {year} {1965})}\BibitemShut {NoStop}%
\bibitem [{\citenamefont {Bezanson}\ \emph {et~al.}(2017)\citenamefont {Bezanson}, \citenamefont {Edelman}, \citenamefont {Karpinski},\ and\ \citenamefont {Shah}}]{Bezanson2017julia}%
  \BibitemOpen
  \bibfield  {author} {\bibinfo {author} {\bibfnamefont {J.}~\bibnamefont {Bezanson}}, \bibinfo {author} {\bibfnamefont {A.}~\bibnamefont {Edelman}}, \bibinfo {author} {\bibfnamefont {S.}~\bibnamefont {Karpinski}},\ and\ \bibinfo {author} {\bibfnamefont {V.~B.}\ \bibnamefont {Shah}},\ }\bibfield  {title} {\bibinfo {title} {Julia: A fresh approach to numerical computing},\ }\href {https://doi.org/10.1137/141000671} {\bibfield  {journal} {\bibinfo  {journal} {SIAM review}\ }\textbf {\bibinfo {volume} {59}},\ \bibinfo {pages} {65} (\bibinfo {year} {2017})}\BibitemShut {NoStop}%
\bibitem [{\citenamefont {Anderson}\ \emph {et~al.}(1999)\citenamefont {Anderson}, \citenamefont {Bai}, \citenamefont {Bischof}, \citenamefont {Blackford}, \citenamefont {Demmel}, \citenamefont {Dongarra}, \citenamefont {Du~Croz}, \citenamefont {Greenbaum}, \citenamefont {Hammarling}, \citenamefont {McKenney},\ and\ \citenamefont {Sorensen}}]{lapack99}%
  \BibitemOpen
  \bibfield  {author} {\bibinfo {author} {\bibfnamefont {E.}~\bibnamefont {Anderson}}, \bibinfo {author} {\bibfnamefont {Z.}~\bibnamefont {Bai}}, \bibinfo {author} {\bibfnamefont {C.}~\bibnamefont {Bischof}}, \bibinfo {author} {\bibfnamefont {S.}~\bibnamefont {Blackford}}, \bibinfo {author} {\bibfnamefont {J.}~\bibnamefont {Demmel}}, \bibinfo {author} {\bibfnamefont {J.}~\bibnamefont {Dongarra}}, \bibinfo {author} {\bibfnamefont {J.}~\bibnamefont {Du~Croz}}, \bibinfo {author} {\bibfnamefont {A.}~\bibnamefont {Greenbaum}}, \bibinfo {author} {\bibfnamefont {S.}~\bibnamefont {Hammarling}}, \bibinfo {author} {\bibfnamefont {A.}~\bibnamefont {McKenney}},\ and\ \bibinfo {author} {\bibfnamefont {D.}~\bibnamefont {Sorensen}},\ }\href@noop {} {\emph {\bibinfo {title} {{LAPACK} Users' Guide}}},\ \bibinfo {edition} {3rd}\ ed.\ (\bibinfo  {publisher} {Society for Industrial and Applied Mathematics},\ \bibinfo {address} {Philadelphia, PA},\ \bibinfo {year} {1999})\BibitemShut {NoStop}%
\bibitem [{\citenamefont {Busch}\ and\ \citenamefont {Anglin}(2000)}]{Busch2000}%
  \BibitemOpen
  \bibfield  {author} {\bibinfo {author} {\bibfnamefont {{\relax Th}.}~\bibnamefont {Busch}}\ and\ \bibinfo {author} {\bibfnamefont {J.~R.}\ \bibnamefont {Anglin}},\ }\bibfield  {title} {\bibinfo {title} {Motion of {{Dark Solitons}} in {{Trapped Bose-Einstein Condensates}}},\ }\href {https://doi.org/10.1103/PhysRevLett.84.2298} {\bibfield  {journal} {\bibinfo  {journal} {Phys. Rev. Lett.}\ }\textbf {\bibinfo {volume} {84}},\ \bibinfo {pages} {2298} (\bibinfo {year} {2000})},\ \Eprint {https://arxiv.org/abs/11018869} {11018869} \BibitemShut {NoStop}%
\bibitem [{\citenamefont {Kuznetsov}\ and\ \citenamefont {Turitsyn}(1988)}]{Kuznetsov1988}%
  \BibitemOpen
  \bibfield  {author} {\bibinfo {author} {\bibfnamefont {E.~A.}\ \bibnamefont {Kuznetsov}}\ and\ \bibinfo {author} {\bibfnamefont {S.~K.}\ \bibnamefont {Turitsyn}},\ }\bibfield  {title} {\bibinfo {title} {Instability and collapse of solitons in media with a defocusing nonlinearity},\ }\href@noop {} {\bibfield  {journal} {\bibinfo  {journal} {Sov. Phys. JETP}\ }\textbf {\bibinfo {volume} {67}},\ \bibinfo {pages} {1583} (\bibinfo {year} {1988})}\BibitemShut {NoStop}%
\bibitem [{\citenamefont {Muryshev}\ \emph {et~al.}(1999)\citenamefont {Muryshev}, \citenamefont {{van Linden van den Heuvell}},\ and\ \citenamefont {Shlyapnikov}}]{Muryshev1999}%
  \BibitemOpen
  \bibfield  {author} {\bibinfo {author} {\bibfnamefont {A.~E.}\ \bibnamefont {Muryshev}}, \bibinfo {author} {\bibfnamefont {H.~B.}\ \bibnamefont {{van Linden van den Heuvell}}},\ and\ \bibinfo {author} {\bibfnamefont {G.~V.}\ \bibnamefont {Shlyapnikov}},\ }\bibfield  {title} {\bibinfo {title} {Stability of standing matter waves in a trap},\ }\href {https://doi.org/10.1103/PhysRevA.60.R2665} {\bibfield  {journal} {\bibinfo  {journal} {Phys. Rev. A}\ }\textbf {\bibinfo {volume} {60}},\ \bibinfo {pages} {R2665} (\bibinfo {year} {1999})}\BibitemShut {NoStop}%
\bibitem [{\citenamefont {Brand}\ and\ \citenamefont {Reinhardt}(2002)}]{Brand2002}%
  \BibitemOpen
  \bibfield  {author} {\bibinfo {author} {\bibfnamefont {J.}~\bibnamefont {Brand}}\ and\ \bibinfo {author} {\bibfnamefont {W.~P.}\ \bibnamefont {Reinhardt}},\ }\bibfield  {title} {\bibinfo {title} {Solitonic vortices and the fundamental modes of the "snake instability": {{Possibility}} of observation in the gaseous {{Bose-Einstein}} condensate},\ }\href {https://doi.org/10.1103/PhysRevA.65.043612} {\bibfield  {journal} {\bibinfo  {journal} {Phys. Rev. A}\ }\textbf {\bibinfo {volume} {65}},\ \bibinfo {pages} {043612} (\bibinfo {year} {2002})}\BibitemShut {NoStop}%
\bibitem [{\citenamefont {Cetoli}\ \emph {et~al.}(2013)\citenamefont {Cetoli}, \citenamefont {Brand}, \citenamefont {Scott}, \citenamefont {Dalfovo},\ and\ \citenamefont {Pitaevskii}}]{cetoliSnakeInstabilityDark2013}%
  \BibitemOpen
  \bibfield  {author} {\bibinfo {author} {\bibfnamefont {A.}~\bibnamefont {Cetoli}}, \bibinfo {author} {\bibfnamefont {J.}~\bibnamefont {Brand}}, \bibinfo {author} {\bibfnamefont {R.~G.}\ \bibnamefont {Scott}}, \bibinfo {author} {\bibfnamefont {F.}~\bibnamefont {Dalfovo}},\ and\ \bibinfo {author} {\bibfnamefont {L.~P.}\ \bibnamefont {Pitaevskii}},\ }\bibfield  {title} {\bibinfo {title} {Snake instability of dark solitons in fermionic superfluids},\ }\href {https://doi.org/10.1103/PhysRevA.88.043639} {\bibfield  {journal} {\bibinfo  {journal} {Phys. Rev. A}\ }\textbf {\bibinfo {volume} {88}},\ \bibinfo {pages} {043639} (\bibinfo {year} {2013})},\ \Eprint {https://arxiv.org/abs/1307.3717} {arxiv:1307.3717} \BibitemShut {NoStop}%
\bibitem [{\citenamefont {Kamchatnov}\ and\ \citenamefont {Pitaevskii}(2008)}]{Kamchatnov2008}%
  \BibitemOpen
  \bibfield  {author} {\bibinfo {author} {\bibfnamefont {A.}~\bibnamefont {Kamchatnov}}\ and\ \bibinfo {author} {\bibfnamefont {L.}~\bibnamefont {Pitaevskii}},\ }\bibfield  {title} {\bibinfo {title} {Stabilization of {{Solitons Generated}} by a {{Supersonic Flow}} of {{Bose-Einstein Condensate Past}} an {{Obstacle}}},\ }\href {https://doi.org/10.1103/PhysRevLett.100.160402} {\bibfield  {journal} {\bibinfo  {journal} {Phys. Rev. Lett.}\ }\textbf {\bibinfo {volume} {100}},\ \bibinfo {pages} {160402} (\bibinfo {year} {2008})}\BibitemShut {NoStop}%
\bibitem [{\citenamefont {Mueller}(2002)}]{Mueller2002b}%
  \BibitemOpen
  \bibfield  {author} {\bibinfo {author} {\bibfnamefont {E.~J.}\ \bibnamefont {Mueller}},\ }\bibfield  {title} {\bibinfo {title} {Superfluidity and mean-field energy loops: {{Hysteretic}} behavior in {{Bose-Einstein}} condensates},\ }\href {https://doi.org/10.1103/PhysRevA.66.063603} {\bibfield  {journal} {\bibinfo  {journal} {Phys. Rev. A}\ }\textbf {\bibinfo {volume} {66}},\ \bibinfo {pages} {063603} (\bibinfo {year} {2002})}\BibitemShut {NoStop}%
\bibitem [{\citenamefont {Mu{\~n}oz~Mateo}\ \emph {et~al.}(2019)\citenamefont {Mu{\~n}oz~Mateo}, \citenamefont {Delgado}, \citenamefont {Guilleumas}, \citenamefont {Mayol},\ and\ \citenamefont {Brand}}]{munozmateoNonlinearWavesBoseEinstein2019a}%
  \BibitemOpen
  \bibfield  {author} {\bibinfo {author} {\bibfnamefont {A.}~\bibnamefont {Mu{\~n}oz~Mateo}}, \bibinfo {author} {\bibfnamefont {V.}~\bibnamefont {Delgado}}, \bibinfo {author} {\bibfnamefont {M.}~\bibnamefont {Guilleumas}}, \bibinfo {author} {\bibfnamefont {R.}~\bibnamefont {Mayol}},\ and\ \bibinfo {author} {\bibfnamefont {J.}~\bibnamefont {Brand}},\ }\bibfield  {title} {\bibinfo {title} {Nonlinear waves of {{Bose-Einstein}} condensates in rotating ring-lattice potentials},\ }\href {https://doi.org/10.1103/PhysRevA.99.023630} {\bibfield  {journal} {\bibinfo  {journal} {Phys. Rev. A}\ }\textbf {\bibinfo {volume} {99}},\ \bibinfo {pages} {023630} (\bibinfo {year} {2019})}\BibitemShut {NoStop}%
\bibitem [{\citenamefont {Komineas}\ and\ \citenamefont {Papanicolaou}(2003)}]{Komineas2003}%
  \BibitemOpen
  \bibfield  {author} {\bibinfo {author} {\bibfnamefont {S.}~\bibnamefont {Komineas}}\ and\ \bibinfo {author} {\bibfnamefont {N.}~\bibnamefont {Papanicolaou}},\ }\bibfield  {title} {\bibinfo {title} {Solitons, solitonic vortices, and vortex rings in a confined {{Bose-Einstein}} condensate},\ }\href {https://doi.org/10.1103/PhysRevA.68.043617} {\bibfield  {journal} {\bibinfo  {journal} {Phys. Rev. A}\ }\textbf {\bibinfo {volume} {68}},\ \bibinfo {pages} {043617} (\bibinfo {year} {2003})}\BibitemShut {NoStop}%
\bibitem [{Lan(1965)}]{Landau1965}%
  \BibitemOpen
  \bibfield  {title} {\bibinfo {title} {46 - {T}he theory of superfluidity of helium {II}},\ }in\ \href {https://doi.org/https://doi.org/10.1016/B978-0-08-010586-4.50051-1} {\emph {\bibinfo {booktitle} {Collected Papers of L.D. Landau}}},\ \bibinfo {editor} {edited by\ \bibinfo {editor} {\bibfnamefont {D.}~\bibnamefont {{ter Haar}}}}\ (\bibinfo  {publisher} {Pergamon},\ \bibinfo {year} {1965})\ pp.\ \bibinfo {pages} {301--330}\BibitemShut {NoStop}%
\end{thebibliography}%

\end{document}